\documentclass[11pt]{article}

\usepackage{amsmath}
\usepackage{amssymb}
\usepackage{graphicx}
\usepackage[T2A,T1]{fontenc}
\usepackage{jheppub}

\hypersetup{
  colorlinks=true,
  citecolor=blue,
  linkcolor=black,
  urlcolor=blue,
  linktocpage,
  linktoc=all
}

\numberwithin{equation}{section}

\newcommand{\lr}[1]{\left(#1\right)}
\def\q{{\mathfrak{q}}}
\def\w{{\mathfrak{w}}}
\def\GB{{\rm GB}}
\def\lgb{\lambda_{\GB}}
\def\coeff#1#2{{\textstyle {\frac {#1}{#2}}}}
\let\hb\hbar 

\DeclareRobustCommand{\dje}{%
  \ifmmode
    \text{{\fontencoding{T2A}\selectfont\cyrdje}}%
  \else
    {\fontencoding{T2A}\selectfont\cyrdje}%
  \fi
}
\renewcommand{\hbar}{\dje} 

\begin{document}

\title{Instability of 5D Gauss-Bonnet black branes}
\author[a]{Alex Buchel,}
\author[b]{Raphael E. Hoult}
\author[b,c]{and Pavel Kovtun}
\affiliation[a]{
{\it \small Department of Physics and Astronomy, University of Western Ontario}\\
{\it \small London, Ontario N6A 5B7, Canada}
}
\affiliation[b]{
{\it\small Department of Physics \& Astronomy, University of Victoria}\\
{\it\small PO Box 1700 STN CSC, Victoria, BC, V8W 2Y2, Canada}
}
\affiliation[c]{
{\it\small School of Mathematical Sciences \& STAG Research Centre}\\
{\it\small University of Southampton, SO17 1BJ, Southampton, United Kingdom}\\
}
\abstract{
We show that Gauss-Bonnet black branes in five-dimensional anti-de Sitter gravity are unstable when the Gauss-Bonnet coupling falls outside the range allowed by the conformal collider bounds.
The unstable modes and the boundary causality violating modes are connected by a phase rotation of complex boundary momentum. 
}

\maketitle

\section{Introduction}
\label{sec:introduction}
The holographic gauge theory-string theory correspondence~\cite{Aharony:1999ti} has been a fruitful research field and a source of inspiration in theoretical physics since the beginning of this century. One of the attractions of the correspondence lies in its use as a tool to learn about (classical or quantum) gravity from quantum field theory and vice versa~\cite{Ammon:2015wua}. Viewing gravity as an effective field theory~\cite{Burgess:2003jk} embedded in a larger string-theoretic framework, quantum corrections to Einstein's general relativity due to massive degrees of freedom arise as higher-derivative corrections to the Einstein-Hilbert action~\cite{Birrell:1982ix}. Such higher-derivative truncated gravitational theories may then in principle be studied in their own right as classical field theories, however caution is warranted. To quote from the textbook of Green, Schwarz and Witten, in string theory ``we cannot really construct a low-energy effective action for the massless fields. What we can do is to study string S-matrix elements and simply construct a classical action for the massless fields that reproduces them''~\cite{Green:1987mn}. Put differently, truncated higher-derivative gravitational theories may not end up healthy; after all, the effective action constructed to reproduce low-energy graviton amplitudes is not meant to be used to describe classical gravitational dynamics as an initial-value problem for partial differential equations. In this paper, we will be interested in an example of such a truncated higher-derivative theory, Gauss-Bonnet (GB) gravity in five dimensions.

We are interested in the GB theory in the context of holography. The action of the theory is a modification of Einstein's action by three curvature-squared terms, where the relative coefficients of the three terms are tuned so that the equations of motion contain no derivatives of the metric higher than second order. The action we consider is
\begin{align}
\label{eq:SGB}
  S = \frac{1}{2 \ell_P^3} \int\!\! d^5x \sqrt{-g} \biggl[ R + \frac{12}{L^2} + \frac{\lgb}{2} L^2 \lr{R^2 - 4 R_{MN} R^{MN} + R_{MNJK} R^{MNJK}} \biggr] \,,
\end{align}
where $\ell_P^3$ is proportional to the five-dimensional Newton's constant, $L$ determines the scale of the (negative) cosmological constant, $R$ is the Ricci scalar, $R_{MN}$ is the Ricci tensor, and $R_{MNJK}$ is the Riemann tensor. The coefficient of the GB term is measured in units of $L$, so that $\lgb$ is a dimensionless coefficient. 
The GB theory is an example of a more general Lovelock class of gravitational theories~\cite{Lovelock:1971yv} whose equations of motion are second order. The second-order equations of motion ensure that Lovelock theories do not suffer from the Ostrogradski instability~\cite{Woodard:2015zca}; this fact has generated considerable interest in such theories~\cite{Padmanabhan:2013xyr}. 

While the absence of the Ostrogradski instability may be viewed as a plus from a theoretical perspective, the Ostrogradski instability is not the only sickness that may plague a classical field theory. For example, a classical field theory with second-order equations of motion is not guaranteed to admit a well-posed initial value problem, or the theory may violate causality, or the classical theory may not lead to a sensible quantum theory when perturbatively quantized. The literature on the GB theory and its holographic applications is substantial, see e.g.~\cite{Camanho:2013pda, Grozdanov:2016fkt} for reviews. In what follows, we will limit ourselves to a few comments which touch on our subsequent stability analysis. 

The theory \eqref{eq:SGB} does not admit maximally symmetric solutions when $\lgb > 1/4$. Hence, in holographic applications, where the anti-de Sitter (AdS) space is viewed as the vacuum of a putative dual conformal field theory, the values $\lgb > 1/4$ are excluded. The value $\lgb = 1/4$ is special, and the action can be written in a Chern-Simons form~\cite{Chamseddine:1989nu}.
Here we will only consider $\lgb<1/4$. In this case, the theory \eqref{eq:SGB} admits an AdS$_5$ vacuum, with $R_{MN} = -4 g_{MN}/L_{\rm AdS}^2$, where $L_{\rm AdS}^2 = 2\lgb L^2/(1-\sqrt{1-4\lgb})$. Linearized perturbation equations about this vacuum are hyperbolic, and the characteristics are on the metric light cone, just like in Einstein gravity, see e.g.~\cite{Devecioglu:2024qiu}. For perturbations of other backgrounds in theory~\eqref{eq:SGB}, things differ substantially from Einstein gravity both in terms of hyperbolicity and superluminality, see e.g.~\cite{Reall:2014pwa}.

The equations of motion of the GB theory are second-order, however they are (unlike Einstein's equations) not linear in the second derivatives of the metric. When one studies perturbations of a given background, the properties of the linearized perturbation equations depend on the background. For some backgrounds, the perturbation equations may be hyperbolic and may propagate signals along metric light cones, like in the Einstein theory. For other backgrounds, the perturbation equations may propagate signals superluminally (i.e.\ outside the metric light cones), or it may happen that the perturbation equations themselves are not even hyperbolic. As hyperbolicity and subluminality are local notions, they may be respected by the perturbation equations in some regions of the background spacetime but not in others. In the region where perturbation equations are hyperbolic, signal propagation may be subluminal in some regions, and superluminal in others. 

For black holes with horizon radius $r_h$ in theory~\eqref{eq:SGB}, perturbation equations are hyperbolic everywhere outside the horizon only if inequalities of the type $F(r_h/L, \lgb)>0$ are satisfied, where the form of $F$ depends on the polarization of the perturbation~\cite{Reall:2014pwa, Andrade:2016yzc}. For black branes ($r_h/L \to \infty$), perturbation equations are not everywhere hyperbolic for $|\lgb|>1/8$~\cite{Brigante:2007nu, Brigante:2008gz,  Buchel:2009tt}. 
In the analysis of quasinormal modes, the violation of hyperbolicity for black hole or black brane perturbations is manifested as a short-wavelength instability, namely quasinormal frequencies with a positive imaginary part at real wave vector when $\lgb$ falls outside the hyperbolic range~\cite{Buchel:2009tt, Konoplya:2017ymp}. In particular, asymptotically AdS$_5$ black branes in theory~\eqref{eq:SGB} are unstable for $|\lgb|>1/8$, suggesting that
\begin{align}
\label{eq:hcl}
  -\frac18 < \lgb < \frac18 
\end{align}
is a necessary consistency condition for the holographic black brane setup.%
\footnote{
The unstable modes at $\lgb=1/4$ can be found analytically~\cite{Grozdanov:2016fkt}.
}

Perturbation equations propagate signals on the metric lightcone at asymptotic AdS infinity. However, in the bulk of the black hole/brane background, effective lightcones for gravitational perturbations may lie outside (or inside) the metric lightcone, depending on the region and polarization of the perturbation~\cite{Brigante:2007nu, Brigante:2008gz,  Buchel:2009tt, Reall:2014pwa}. The insight of \cite{Brigante:2008gz} was to realize that a signal may travel from the AdS boundary into the bulk and back to the boundary on an effective superluminal bulk lightcone, producing superluminal propagation on the boundary. For black branes, the conclusion is that the theory \eqref{eq:SGB}, when taken on its own, (i.e.\ without any other fields or higher-derivative corrections), can not describe a putative boundary conformal field theory unless boundary superluminality is excluded, which can happen only for $-7/36<\lgb<9/100$~\cite{Brigante:2007nu, Brigante:2008gz, Buchel:2009tt}. 

One can also consider perturbations of a (singular) shock-wave background~\cite{Camanho:2014apa}.
Demanding causality in the singular background would eliminate all $O(1)$ values of $\lgb$ (unless other degrees of freedom are included)~\cite{Camanho:2014apa}. Causality constraints have also been studied in smooth black-hole backgrounds rather than singular shockwave geometries \cite{Papallo:2015rna}. While time advances persist in suitable regimes, the interpretation of these effects and their relation to macroscopic causality violations becomes more subtle.

In holography, the coefficient $\lgb$ of the theory~\eqref{eq:SGB} is related to ``central charge'' coefficients $c$ and $a$ which appear in the conformal anomaly of the dual field theory as~\cite{Nojiri:1999mh, Myers:2010jv}
\begin{align}
  \label{eq:ac}
  & c = \frac{\pi^2}{2^{3/2}}\,\frac{L^3}{\ell_P^3}\,
  (1+\sqrt{1-4\lgb})^{3/2}\,\sqrt{1-4\lgb}\,,\nonumber\\
  & a =\frac{\pi^2}{2^{3/2}}\,\frac{L^3}{\ell_P^3}\,
  (1+\sqrt{1-4\lgb})^{3/2}\,\left(3\sqrt{1-4\lgb}-2\right)\,.
\end{align}
Thus, field-theoretic constraints on the conformal anomaly coefficients $c$ and $a$ can provide constraints on $\lgb$, excluding those values of $\lgb$ for which the theory \eqref{eq:SGB} can not describe any boundary conformal field theory. Field-theoretic constraints on $c$ and $a$ have been identified in~\cite{Hofman:2008ar}, using ``conformal collider'' arguments. Translating the conformal collider constraint $-1/2 < (c-a)/c < 1/2$ in superconformal\footnote{The conformal collider bounds are less stringent for general unitary parity-preserving four-dimensional CFTs than for superconformal theories \cite{Hofman:2008ar,Hofman:2016awc}.} field theories to $\lgb$, one finds
\begin{align}
\label{eq:ccl}
  -\frac{7}{36} < \lgb < \frac{9}{100} \,,
\end{align}
which coincides with the constraints from the absence of boundary superluminality \cite{Hofman:2008ar, Buchel:2009tt}.

As mentioned earlier, black brane backgrounds in the GB theory develop unstable quasinormal modes when hyperbolicity of the perturbation equations is violated, making black branes unstable if \eqref{eq:hcl} does not hold. In this paper we will show that black brane backgrounds also develop unstable quasinormal modes when boundary causality is violated, making black branes unstable if \eqref{eq:ccl} does not hold. In other words, violations of either hyperbolicity or boundary causality give rise to unstable quasinormal modes. 
Both of these instabilities are short-wavelength instabilities, however their mechanisms are quite distinct. 

A simple example to keep in mind is the telegraph equation $(\partial_t^2 + \partial_t - c_s^2 \partial_x^2)\phi(t,x) = 0$, which describes damped signals propagating with speed $c_s$. The corresponding frequency $\omega$ and the wavevector $k$ satisfy $\omega^2 + i \omega - c_s^2 k^2 = 0$. Unstable modes correspond to solutions $(\omega, k)$ such that ${\rm Im}(\omega) > 0$ at real $k$. Violation of hyperbolicity corresponds to $c_s^2<0$, which gives rise to an unstable mode. Violation of causality corresponds to $c_s^2>1$; in this case the modes are naively stable, however in a Lorentz-boosted reference frame, one finds a mode with ${\rm Im}(\omega) > 0$ at real $k$ when the boost velocity $v>1/c_s$. The instability is a reflection of the special-relativistic statement that the time-ordering of two events is reversed when $v>1/c_s$~\cite{Tolman-1917}, so what looks like dissipation in one reference frame looks like an instability in another. See e.g.~\cite{Gavassino:2021owo} for a recent discussion. We will see that boosted black branes in the GB theory exhibit a similar instability when $v$ exceeds a critical value. The instability we identify is the complementary view of the boundary causality violation discussed in~\cite{Brigante:2007nu, Brigante:2008gz, Buchel:2009tt}.

In practice, working with a boosted state may be inconvenient. However, the instability of a boosted state is also visible in the spectrum of the state at rest, albeit at complex wavevector. Indeed, suppose that the state at rest happens to have an eigenmode $(\omega, k)$ with ${\rm Im}(\omega) > 0$ and ${\rm Im}(\omega)^2 > {\rm Im}(k)^2$. In a boosted state with boost velocity $v$, one has
\begin{align}
 (\omega', k') = \frac{1}{\sqrt{1-v^2}} \left( \omega + k v, k + \omega v\right) \,.
\end{align}
Thus, one can choose a value of $v$ with $0<v^2<1$ which sets ${\rm Im}(k') = 0$. At that value of $v$, 
$$
  {\rm Im}(k') = 0\,,\ \ \ \ {\rm Im}(\omega') = \sqrt{1-v^2} \;{\rm Im}(\omega) >0 \,,
$$
indicating that the boosted state is unstable~\cite{Gavassino:2023myj}. 

As an example, suppose we are looking for an instability of the boosted state with $(\omega',k') = (i\gamma, 0)$, with $\gamma>0$. Then the corresponding mode in the state at rest,
\begin{align}
  (\omega, k) = \left( \frac{i\gamma}{\sqrt{1-v^2}}, \frac{-i\gamma v}{\sqrt{1-v^2}} \right) \,,
\end{align}
has purely imaginary $\omega$ and $k$, with $\omega^2/k^2 = 1/v^2 > 1$. Conversely, if the state at rest has an eigenmode with purely imaginary $\omega$ and $k$ such that ${\rm Im}(\omega)>0$ and $u^2 \equiv \omega^2/k^2 >1$, then performing a boost with velocity $1/u$ (with the corresponding sign of $u$) gives rise to a mode in the boosted reference frame whose frequency is positive imaginary at zero wave vector.
We will show both analytically and numerically that such unstable modes indeed exist for GB black branes, unless $\lgb$ is constrained to lie in the range \eqref{eq:ccl} allowed by the conformal collider bounds of the putative dual field theory. More generally, we will find that GB black branes have modes whose boundary dispersion relations violate the condition
\begin{align}
\label{linstab}
 {\rm Im}\, \omega \leq |{\rm Im} \, k| \,,
\end{align}
which is not allowed in a relativistic theory; see~\cite{Heller:2022ejw} for a recent discussion in the context of transport coefficients.

We will use the following terminology. Following ref.~\cite{Gavassino:2023myj}, we will call eq.~\eqref{linstab} the covariant stability condition. We will refer to the modes with large real $k$, with $|{\rm Re}(\omega) /k| >1$, whose existence was suggested in~\cite{Brigante:2007nu, Brigante:2008gz, Buchel:2009tt}, as causality-violating modes. We will refer to the modes with imaginary $k$ and imaginary $\omega$, with $|\omega/k| >1$ as unstable modes.  

The remainder of the paper proceeds as follows. In Section~\ref{sec:analytics}, we will show that static black branes in the Gauss-Bonnet theory \eqref{eq:SGB} have quasinormal modes with positive imaginary $\omega$ at imaginary $k$, such that $\omega^2/k^2$ approaches a constant greater than 1 at large $|k|$. In the putative dual field theory, such modes would give rise to poles in the retarded functions of the energy-momentum tensor at the values of $\omega$ and $k$ which violate \eqref{linstab}. By the above argument, black branes boosted along a translation-invariant boundary direction must then have modes with positive imaginary $\omega$ at vanishing $k$, indicating that boosted black branes are unstable, if $\lgb$ falls outside the range~\eqref{eq:ccl}. We confirm these findings numerically in Section~\ref{sec:numerics}, and discuss our results in Section~\ref{sec:discussion}.

\section{Unstable quasinormal modes}
\label{sec:analytics}

\subsection{Black brane background}
\label{sec:background}
We start with the effective action \eqref{eq:SGB}. We find it convenient to parameterize $\lgb$ as
\begin{equation}
   \beta\equiv \coeff12 +\coeff12 \sqrt{1-4\lgb}\,, \qquad \lgb=\beta-\beta^2\,,
\label{defb2}
\end{equation}
so that $\beta > 1/2$, and Einstein gravity corresponds to $\beta = 1$. 
The condition \eqref{eq:ccl} (reflecting the absence of superluminal propagation on the boundary, or equivalently, the conformal collider bounds on $c$ and $a$) then translates into 
\begin{equation}
  \frac{9}{10}\le \beta\le \frac 76 \,.
\label{uvcausal}
\end{equation}
In what follows, without the loss of generality, we set $L=1$.
Varying the action~\eqref{eq:SGB} along with appropriate counterterms yields the equations of motion
\begin{align}
\label{eq:GB-eom}
    R_{MN} - \frac{1}{2} R g_{MN} - 6 g_{MN} & = \frac{\lgb}{2} \biggl[ \frac{1}{2} g_{MN} \lr{R^2 - 4 R_{MN} R^{MN} + R_{MNJK} R^{MNJK}} \nonumber\\
    & {-} 2 R R_{MN} + 4 R_{ML}R^L_{\,\,N} 
    + 4 R^{KL}R_{KMLN} - 2 R_{MJKL}R_N^{\,\,\,\,JKL} \biggr] .
\end{align}
To study thermal equilibrium states of the model \eqref{eq:SGB}, we use the bulk metric ansatz
\begin{equation}
  \label{metric}
  ds^2= \frac{r_h^2}{x}\left(-f_1 \beta\ dt^2+\sum_{i=1}^3 dx_i^2\right)+\frac{1}{f_2} \frac{dx^2}{4x^2}\,,
\end{equation}
where the metric warp factors $f_1$, $f_2$  are functions of the radial coordinate $x$ only, $x\in (0,1)$.
Solving \eqref{eq:GB-eom} gives 
\begin{equation}
  f_1=f_2\equiv f(x)=\frac{1-\sqrt{1-4(\beta-\beta^2)(1-x^2)}}{2(\beta-\beta^2)}\,,
\label{cft}
\end{equation}
with $r_h$ in \eqref{metric} as an integration constant. 
The coordinate $x$ is related to the often-used AdS$_5$ radial coordinate $r$ via $x = r_h^2/r^2$, where $r=r_h$ is the horizon. The temperature of the background is $T = r_h \sqrt{\beta}/\pi$.
Thermodynamics of the solution \eqref{metric} is discussed in Appendix~\ref{app:thermodynamics}. 

Applying a Lorentz transformation to the boundary coordinates $x^\mu = (t, x^i)$, we get the black brane solution in the form
\begin{equation}
\label{eq:BBB}
    ds^2 = -\frac{r_h^2 \beta f(x)}{x} u_\mu u_\nu dx^\mu dx^\nu + \frac{dx^2}{4 x f(x)} + \frac{r_h^2}{x} \Delta_{\mu\nu} dx^\mu dx^\nu\,,
\end{equation}
where $u^\mu$ is a constant timelike unit vector (boost 4-velocity), $\Delta_{\mu\nu} \equiv \eta_{\mu\nu} + u_\mu u_\nu$, and $\eta_{\mu\nu}$ is the flat Minkowski metric in 4 dimensions. The boundary energy-momentum tensor corresponding to this geometry has the form $T^{\mu\nu} = 3P(T) u^\mu u^\nu + P(T) \Delta^{\mu\nu}$, where $P(T) \propto T^4$ is the equilibrium pressure of the putative dual conformal field theory on the boundary.

Our goal will be to show that, unless $\beta$ lies in the range~\eqref{uvcausal}, there is always a boundary reference frame (and therefore a boosted equilibrium state) in which the black brane~\eqref{eq:BBB} is unstable to linear perturbations. That said, working directly with the boosted black brane \eqref{eq:BBB} is rather tedious. Instead, we will take a different approach, and work with perturbations about the black brane~\eqref{metric}. Information about the stability and causality of a thermal system in a boosted reference frame may be determined by an analysis of perturbations in the rest frame at complex wavevector.

\subsection{Perturbation equations}
We consider linearized perturbations of the metric~\eqref{metric} of the form $h_{MN} = e^{-i \omega t + i k z}\xi_{MN}(x)$. Using the isotropy of the equilibrium solution~\eqref{metric}, the wave-vector is chosen along $z\equiv x_3$. We will be primarily working with the dimensionless 
\begin{equation}
    \w = \frac{\omega}{2\pi T}, \qquad \q = \frac{k}{2\pi T}\,.
\end{equation}
Rotation symmetry of the equilibrium background~\eqref{metric} implies that the equations of motion~\eqref{eq:GB-eom} separate into three distinct channels upon linearization: a helicity-2 channel (scalar), a helicity-1 channel (shear), and a helicity-0 channel (sound), analogous to what happens for black branes in Einstein gravity~\cite{Kovtun:2005ev}. In all three cases, the linearized equations may be written in terms of variables $Z_a(x)$ which are built from linear combinations of the components of $\xi_{MN}(x)$, and obey equations of the form \cite{Kovtun:2005ev, Buchel:2009tt, Buchel:2009sk} 
\begin{equation}
\label{eq:QNM_master_eq}
    Z_a''(x) + {\cal A}_a(x,\w,\q^2) Z_a'(x) + {\cal B}_a(x,\w,\q^2) Z_a(x) = 0\,.
\end{equation}
The index $a$ denotes the helicity of the perturbation ($a=2$ for scalar, $a=1$ for shear, $a=0$ for sound). There is no summation over ``$a$'' in eq.~\eqref{eq:QNM_master_eq}. The coefficients ${\cal A}_a$ and ${\cal B}_a$ depend on $\q^2$ by rotation invariance. For the shear and sound channels of GB black branes, ${\cal A}_a$ and ${\cal B}_a$ were first derived in~\cite{Buchel:2009tt}; the explicit expressions are given in Appendix~\ref{app:QNM_coeffs}.

The boundary $x=0$ and the horizon $x=1$ are regular singular points of the differential equations~\eqref{eq:QNM_master_eq}. Near the horizon, the indices are $\pm i\w/2$, just like in Einstein gravity \cite{Buchel:2008sh}, and we choose the infalling solution for $Z_a(x)$, whose near-horizon expansion starts with $(1-x)^{-i\w/2}$.  Near the boundary, the indices are $0$ and $2$ just like in the Einstein gravity \cite{Benincasa:2005qc}. Thus, the solution which is infalling near the horizon, will behave near the boundary as
\begin{align}
  Z_a(x) = F_{(a)}(\w, \q^2) (1 + \dots) + G_{(a)}(\w, \q^2) x^2 + \dots\,,
\end{align}
where the dots denote higher powers of $x$. The quasinormal modes are obtained by imposing the Dirichlet condition at the boundary, thus setting
\begin{align}
\label{eq:qnmf}
  F_{(a)}(\w, \q^2) = 0 \,.
\end{align}
Solutions of eq.~\eqref{eq:qnmf} give the singularities of the retarded functions of the energy-momentum tensor in the boundary field theory~\cite{Kovtun:2005ev}.

The differential equation \eqref{eq:QNM_master_eq} can be rewritten in the form of Schr\"odinger's equation. To this end, define a new variable $y(x)$,
\begin{equation}
    \frac{dy}{dx} = g_a(x)\,,
\end{equation}
with $g_a(x)$ sign-definite on $(0,1)$ (so that $y(x)$ is monotonic), and $g_a(x)$ is to be specified momentarily. In the resulting differential equation for $Z_a(y)$, we further take
\begin{align}
  Z_a(y) = \frac{\psi_a(y)}{{\cal C}_a(y)} \,,
\end{align}
and choose ${\cal C}_a(y)$ to eliminate the first derivative in the equation for $\psi_a(y)$,
\begin{align}
\label{eq:CC}
  \frac{{\cal C}_a'(y)}{{\cal C}_a(y)} = \frac12 \left( \frac{g_a'}{g_a^2} + \frac{{\cal A}_a}{g_a} \right) .
\end{align}
We now choose the function $g_a(x)$ so that the resulting differential equation for $\psi_a(y)$, in the limit $|\q^2|\to\infty$, $|\w^2| \to\infty$, with $|\w^2|/|\q^2|$ fixed, would take the form 
\begin{align}
  \psi_a''(y) + \q^2 U_a^{(0)}(y) \psi_a(y) = -\w^2 \psi_a(y) \,,
\end{align}
where the potential $U_a^{(0)}(y)$ is finite in the limit. This is achieved by $g_a(x) = -1/(x^{1/2} f(x))$, the same function in all three symmetry channels~\cite{Buchel:2009sk}. 
The relation between $x$ and $y$ near the boundary and near the horizon is
\begin{align}
  y(x\to0) \sim -2 \beta \, x^{1/2} \,,\ \ \ \ 
  y(x\to1) \sim \coeff12 \ln(1-x) \,.
\end{align}
Given $g_a(x)$, the function ${\cal C}_a$ is obtained from \eqref{eq:CC}. We now define 
\begin{align}
  \hbar^2 \equiv -\frac{1}{\q^2}\,,\ \ \ \ E\equiv - \frac{\w^2}{\q^2}\,.
\end{align}
Perturbation equations~\eqref{eq:QNM_master_eq} then become
\begin{align}
\label{eq:Scheq}
  -\hbar^2 \frac{d^2 \psi_a}{dy^2} + U_a(y) \psi_a(y) = E \psi_a(y) \,,
\end{align}
where the effective potential $U_a(y)$ is determined by the coefficients ${\cal A}_a$ and ${\cal B}_a$ in \eqref{eq:QNM_master_eq}. By construction, $U_a(y)$ stays finite in the limit $\hbar\to0$. 
Before we discuss the boundary conditions on $\psi_a(y)$, let us identify the solutions we'll be looking for.

\subsection{Signatures of the instability}
\label{sec:instab-2}

The quasinormal frequencies of the background \eqref{metric} are determined by eq.~\eqref{eq:qnmf}. Even though it appears that eq.~\eqref{eq:qnmf} only determines the spectrum of the quasinormal modes for the black brane ``at rest'', it in fact determines the quasinormal spectrum for the background \eqref{eq:BBB} as well. 
Define the 4-vector $Q = (\w, 0, 0, \q)$, whose Lorentz transformation in the $z$-direction is $Q' = (\w', 0, 0, \q') = (\w + \q v, 0, 0, \q + \w v)/\sqrt{1-v^2}$. 
Then in a state characterized by the 4-velocity $u^\mu = (1, 0, 0, v)/\sqrt{1-v^2}$, the spectrum is determined by 
\begin{align}
\label{eq:FF}
  F_{(a)}(w,z) = 0\,,
\end{align}
where $w \equiv - Q^\mu \eta_{\mu\nu} u^\mu$, $z\equiv Q^\mu (\eta_{\mu\nu} + u_\mu u_\nu) Q^\nu$. These reduce to $w=\w$ and $z=\q^2$ when $v=0$, and more generally $w=\w'$, $z = \q'^2$. Thus, if a given pair $(\w, \q^2)$ is a solution to \eqref{eq:qnmf}, representing a quasinormal mode in the state ``at rest'', then the boosted pair $(\w', \q'^2)$ is a solution to \eqref{eq:FF}, representing a quasinormal mode in the boosted state.

We will be looking for quasinormal modes of \eqref{metric} which satisfy ${\rm Im}(\w) > |{\rm Im}(\q)|$. We will consider modes $(\w, \q)$ which develop unstable gaps after a boost, namely $(\w', \q') = (i\Gamma, 0)$, with $\Gamma>0$. Such modes are given by
\begin{align}
\label{eq:wq-2}
  (\w, \q) = \frac{i\Gamma}{\sqrt{1-v^2}} \left( 1, -v \right) ,
\end{align}
with both $\w$ and $\q$ purely imaginary, $\w^2/\q^2 = 1/v^2 > 1$, and hence for the unstable modes
\begin{equation}
\label{eq:hbarE}
    \hbar^2 =  \frac{(1-v^2)}{v^2 \Gamma^2}>0 \,, \qquad E = -\frac{1}{v^2} < -1\,.
\end{equation}
For the mode~\eqref{eq:wq-2} with $\Gamma>0$, the boundary condition on $Z_a(x)$ at the horizon becomes
\begin{equation}
    Z_a(x) \propto (1-x)^{-i\w/2} = (1-x)^{\frac{\Gamma}{2\sqrt{1-v^2}}}\,,
\end{equation}
thus $Z_a(x)$ obeys Dirichlet conditions at both the horizon $x=1$ and the boundary $x=0$. 
We will later show that in all three channels,
\begin{equation}
\label{eq:C_asymptotics}
    {\cal C}_a(x) = 
    \begin{cases}
        O(x^{-3/4}),& x\to0\,,\\[5pt]
        O(1)\,, & x \to 1 \,.
    \end{cases}
\end{equation}
As $Z_a(x) = O(x^2)$ at the boundary, $\psi_a(x)$ also obeys Dirichlet boundary conditions at both $x=0$ and $x=1$. The question of finding the modes described above is therefore mapped directly onto a one-dimensional quantum mechanical problem for a potential with infinite walls at $x=0$ and $x=1$ (or, rather, at $y=0$ and $y=-\infty$). 
In the quantum mechanical problem, the unstable modes correspond to bound states with energies $E<-1$. We now discuss how the above considerations play out in all three symmetry channels.

\subsection{Helicity-2 perturbations (scalar channel)}
\label{subsec:scalar_analytics}
The coefficients ${\cal A}_2(x)$ and ${\cal B}_2(x)$ of eq.~\eqref{eq:QNM_master_eq} are given in Appendix~\ref{app:QNM_coeffs_scalar}. Further, eq.~\eqref{eq:CC} may be directly integrated to find the scaling factor ${\cal C}_2(x)$,
\begin{equation}
    {\cal C}_2(x) = x^{-3/4} \lr{4 \beta x^2 (1{-}\beta) + (2 \beta{-}1)^2}^{-1/4} \,,
\end{equation}
which does indeed have the behaviour~\eqref{eq:C_asymptotics} near the horizon and the boundary. The Schr\"odinger equation for $\psi(y)\equiv \psi_2(y)$ may be written in the form
\begin{equation}
\label{eq:Schrodinger-scalar}
    -\hbar^2 \psi''(y) + \left( U_2^{(0)}(x) + \hbar^2 U_2^{(1)}(x) \right) \psi(y) = E \psi(y)\,,
\end{equation}
where $U_2^{(0)}(x)$ and $U_2^{(1)}(x)$ do not depend on $\hbar^2$, and are given by%
\footnote{
  Recall that $\hbar^2 = -1/\q^2 >0$, so our $U_2^{(1)}$ is defined as $U_2 = U_2^{(0)} - (1/\q^2) U_2^{(1)}$.
}
\begin{subequations}
    \begin{align}
        U_2^{(0)}(x) &= -\beta f \biggl(1+8 (1-\beta)\beta - 12 (1-\beta)\beta f (1- (1-\beta)\beta f)\biggr)\\
        &\times \biggl(1-2 (1-\beta) \beta f\biggr)^{-2}\,,\nonumber\\
        U_2^{(1)}(x) &= \frac{1}{16} f \biggl(x\left(2 f \beta (\beta-1)+1\right)^4\biggr)^{-1}\
\biggl(240 \beta^4 (\beta-1)^4 f^5+320 \beta^3 (\beta-1)^3 f^4 \nonumber \\
&-8 \beta^2 (24 \beta^2-24 \beta+1) (\beta-1)^2 f^3+16 \beta (\beta-1) (2 \beta^2
-2 \beta-7) f^2\nonumber\\
&+(192 \beta^4-384 \beta^3+464 \beta^2-272 \beta-9) f-64 \beta (\beta-1)+24\biggr) .
    \end{align}
\end{subequations}
In the potential in eq.~\eqref{eq:Schrodinger-scalar}, $x$ is to be expressed as a function of $y$. 
In order to identify the unstable quasinormal modes of the type described in Sec.~\ref{sec:instab-2}, we need to find bound states in the potential $U_2 = U_2^{(0)} + \hbar^2 U_2^{(1)}$, with $\psi(x)$ vanishing at $x=0$ ($y=0$) and $x=1$ ($y=-\infty$). The ``energy'' $E$ of bound states corresponding to such unstable modes must be such that $E<-1$, according to \eqref{eq:hbarE}. In order to identify such modes, one can proceed as follows: 
\begin{enumerate}
\itemsep0pt
    \item Fix a value for $\hbar$;
    \item Find a bound state energy $E<-1$ by solving~\eqref{eq:Schrodinger-scalar} with Dirichlet boundary conditions (provided such a state exists);
    \item Use that value of $E$ and~\eqref{eq:hbarE} to determine the magnitude of the boost velocity $v$;
    \item Use that $v$ and the known $\hbar$ to determine $\Gamma>0$ from \eqref{eq:hbarE};
    \item Use $\Gamma$ and $v$ to find $\w$ and $\q^2$ from~\eqref{eq:wq-2}.
\end{enumerate}
The ``quantum mechanical'' eigenvalue problem for $\psi$ simplifies in the limit $\hbar \to 0$, which corresponds to either $\Gamma \to\infty$ or $v \to 1^{-}$. In this limit, only $U_2^{(0)}(x)$ contributes to the potential; moreover, as $\hbar\to0$, the ground state energy $E_0$ is simply given by the minimum of the potential.%
\footnote{
In one-dimensional quantum mechanics with Hamiltonian $H$ and potential $V(x)$, the energy of the ground state has to satisfy  $E_{\rm g.s.} \geq V_{\rm min}$, as well as $E_{\rm g.s.} \leq E_\psi \equiv \langle \psi | H | \psi\rangle$, for any normalizable state $\psi$. Choosing $\psi$ as a wavepacket of width $l\to0$ localized at the minimum of $V(x)$ gives kinetic energy contribution to $E_\psi$ as $O(\hb^2/l^2)$, and potential energy contribution to $E_\psi$ as $V_{\rm min}$. Choosing $l$ to scale as $\hb^{1/2}$, one has $E_\psi = V_{\rm min} + O(\hb)$, hence $E_{\rm g.s.} = V_{\rm min}$ as $\hb\to0$.
}
\begin{figure}
\centering
\includegraphics[width=0.6\textwidth]{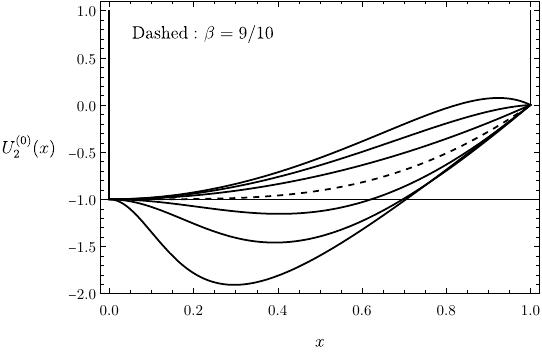}
\caption{
Plots of $U_2^{(0)}(x)$ for $\beta = 0.6, 0.7, 0.8, 0.9, 1.0, 1.1, 1.2$ in the helicity-2 channel. The potential for $\beta = 9/10$ is shown by the dashed line. The potentials for $\beta < 9/10$ are the curves below the dashed line; the potentials for $\beta > 9/10$ are the curves above the dashed line. Vertical lines represent infinite potential walls, reflecting Dirichlet boundary conditions on $\psi$ at $x=0$ and $x=1$. A bound state in this potential with $E<-1$ implies the existence of an unstable gapped mode in the quasinormal spectrum of the boosted black brane (as $\hbar\to0$) at that value of~$\beta$. Only the potentials with $\beta < 9/10$ (or $\lgb>9/100$) have such bound states corresponding to unstable modes. 
}
\label{fig:U10xy}
\end{figure}%
The potential $U_2^{(0)}(x)$ is shown in Figure~\ref{fig:U10xy}; the potential $U_2^{(0)}(y)$ looks qualitatively the same, ``stretched'' over the range of $y$. Near the boundary $x=0$, we have
\begin{equation}
  \label{eq:u20}
    U_2^{(0)}(x) = -1 + \frac{\beta (10 \beta - 9)}{(2 \beta - 1)^2} x^2 + O(x^4)\,.
\end{equation}
As a result, as $\hbar\to0$, the potential $U_2(x)$ dips below $-1$ for $\beta < 9/10$ (corresponding to $\lgb>9/100$), and there will be a bound state with energy $E<-1$. 
Note that our potential $U_2^{(0)}(x)$ is precisely the negative of the effective potential which was used in previous discussions of boundary causality violation in GB gravity, such as~\cite{Brigante:2008gz, Buchel:2009tt}. In those works, it was argued that boundary superluminality is related to a near-boundary bump in their potential; in our case, the same condition in $\lgb$ indicates an instability (in a boosted reference frame) manifested through a near-boundary dip in our potential. 

\begin{figure}
\centering
\includegraphics[width=0.46\textwidth]{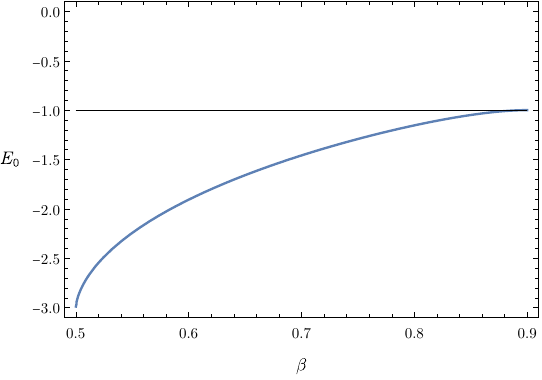}
\hspace{0.05\textwidth}
\includegraphics[width=0.46\textwidth]{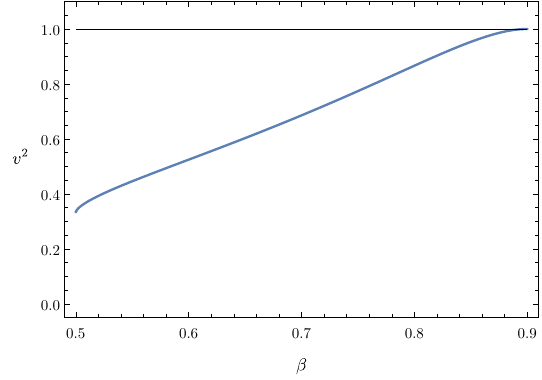}
\caption{
The ground state energy $E_0(\beta)$ in the limit $\hbar\to0$ (left), and the corresponding boost velocity squared $v^2(\beta)$ required to generate an unstable gapped mode at zero wave vector (right) in the scalar channel, plotted in the unstable region $1/2 < \beta < 9/10$. The boost velocity is bounded from below by $v^2(\beta = 1/2) = 1/3$. 
}
\label{fig:hel2_min}
\end{figure}

The minimum value of the potential can be found analytically (as $\hbar\to0$), though the expression for $E_0(\beta) = {\rm min}\, U_2^{(0)}$ is not illuminating, and we will not present it here. As $\beta\to1/2$, the potential $U_2^{(0)}(x) \to 3(x-1)$. As a result, $E_0(\beta)$ in the limit $\hbar\to0$ never dips below $-3$. Near $\beta=1/2$ and $\beta = 9/10$ we have 
\begin{align}
   E_0 = -3 + 2^{1/3} 3^{5/3} \left( \beta - \coeff12 \right)^{2/3} + \dots \,,\ \ \ \ 
   E_0 = -1 - \coeff{625}{22} \left( \beta - \coeff{9}{10} \right)^2 + \dots\,.
\label{eq:E0-scalar}
\end{align}
The dependence of $E_0$ and of the corresponding $v^2 = -1/E_0$ on $\beta$ is shown in Fig.~\ref{fig:hel2_min}.

Our instability discussion of the boosted black branes thus far was done for $\hbar\to 0$, namely with the unstable modes having ${\rm Im}(\w)\gg 1$. The finite-$\hbar$ corrections will not change the reached conclusions, provided $\hbar$ remains sufficiently small. The $O(\hbar^2)$ corrections to $U_2^{(0)}(x)$ are monotonic and positive; in particular, at small $x$ we have 
\begin{align}
  U_2(x) = \frac{15\hbar^2}{16 \beta^2} \frac{1}{x} - 1 + O(\hbar^2 x) \,.
\end{align}
Such finite-$\hbar$ corrections would always lift the dip of the potential $U_2^{(0)}$ in the instability range of the Gauss-Bonnet parameter $1/2<\beta<9/10$, eventually leading to the disappearance of the bound state with $E<-1$, and thus the disappearance of the unstable modes discussed. Likewise, increasing $\hbar$ from zero lifts the minimum of $U_2$, and thus requires larger boost velocity for an instability to exist. We will see these features in explicit computations of the unstable quasinormal modes in Section \ref{sec:numerics}.

Consider now $U_2^{(0)}(x)$ near the horizon, at $x=1$,
\begin{align}
\label{eq:U2h}
 U_2^{(0)}(x)=2\beta (8\beta^2-8\beta-1)\,\, (1-x)+O((1-x)^2)\,.
\end{align}
Since $U_2^{(0)}(x=0)=-1<0$ \eqref{eq:u20}, the potential will necessarily develop
a ``dip'' and a ``bump'' right in front of the horizon provided
\begin{align}
\label{eq:b2hyp}
8\beta^2-8\beta-1 >0\qquad \Longrightarrow\qquad \beta> \frac 12+\frac
{\sqrt 6}{4}\ \Longleftrightarrow\ \lgb<-\frac 18\,, 
\end{align}
see, for example, the potential $U_2^{(0)}$ at $\beta=1.2$ in
Figure~\ref{fig:U10xy}. For the discussion here, this new
near-horizon dip is irrelevant, as it supports
in the limit $\hbar\to 0$ the bound states
with $E>0$, violating \eqref{eq:hbarE}. However,
in the 1D quantum mechanical problem related to the
scalar channel QNMs discussed in
\cite{Brigante:2007nu,Buchel:2009tt,Buchel:2009sk},
the bound states supported near the horizon of $-U_2^{(0)}$,
correspondingly under the "bump'' of \eqref{eq:U2h}, are related to
a different class of unstable QNMs of GB black branes at rest
in the limit of real $\q\to \infty$. These are the
QNM instabilities associated with the violation of the hyperbolicity
of GB black branes for $\lgb<-1/8$ \cite{Buchel:2009tt,Konoplya:2017ymp}.

\subsection{Helicity-1 perturbations (shear channel)}
\label{subsec:shear_analytics}
\begin{figure}
\centering
\includegraphics[width=0.6\textwidth]{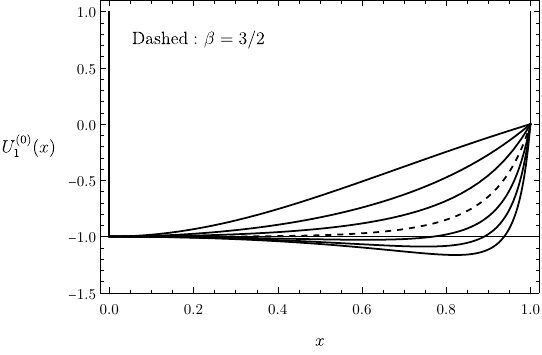}
\caption{
Plots of $U_1^{(0)}(x)$ for $\beta = 0.9, 1.1, 1.3, 1.5, 1.7, 1.9, 2.1$ in the helicity-1 channel. The potential for $\beta = 3/2$ is shown by the dashed line. The potentials for $\beta < 3/2$ correspond to the curves above the dashed line; the potentials for $\beta > 3/2$ correspond to the curves below the dashed line. Vertical lines represent infinite potential walls, reflecting Dirichlet boundary conditions on $\psi$ at $x=0$ and $x=1$. A bound state in this potential with $E<-1$ implies the existence of an unstable gapped mode in the quasinormal spectrum of the boosted black brane (as $\hbar\to0$) at that value of~$\beta$. Only the potentials with $\beta > 3/2$ (or $\lgb<-3/4$) have such bound states corresponding to unstable modes. 
}
\label{fig:U20_analytic}
\end{figure}

\begin{figure}
\centering
\includegraphics[width=0.46\textwidth]{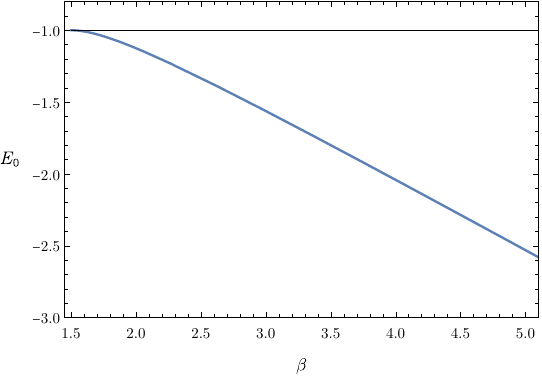}
\hspace{0.05\textwidth}
\includegraphics[width=0.46\textwidth]{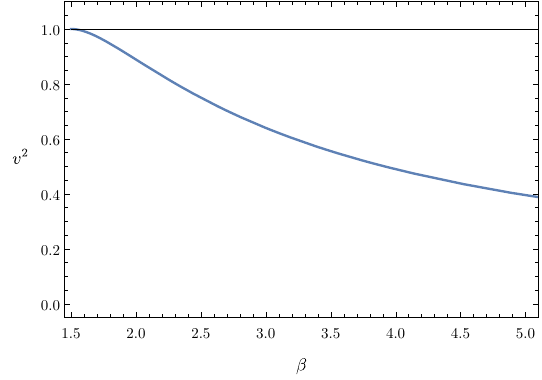}
\caption{
The ground state energy $E_0(\beta)$ in the limit $\hbar\to0$ (left), and the corresponding boost velocity squared $v^2(\beta)$ required to generate an unstable gapped mode at zero wave vector (right) in the shear channel, plotted in the unstable region $\beta > 3/2$. The boost velocity approaches zero at large $\beta$ as $v^2 \sim 2/\beta$.
}
\label{fig:hel2_E0}
\end{figure}

Let us now repeat the analysis of the previous section for the shear channel. The coefficients ${\cal A}_1(x)$, ${\cal B}_1(x)$ of eq.~\eqref{eq:QNM_master_eq} may be found in Appendix~\ref{app:QNM_coeffs_shear}. The integrating factor ${\cal C}_1(x)$ satisfies eq.~\eqref{eq:shear_logC2}, and behaves according to~\eqref{eq:C_asymptotics} near the horizon and boundary. Similar to the scalar channel, eq.~\eqref{eq:QNM_master_eq} may again be written as a Schr\"odinger equation for $\psi(y)\equiv \psi_1(y)$,
\begin{equation}
    - \hbar^2 \psi''(y) + \lr{U_1^{(0)}(x) + \hbar^2 U_1^{(1)}(x)} \psi(y) = E \psi(y)\,.
\end{equation}
The $\hbar$-independent part of the potential $U_1^{(0)}(x)$ is given by
\begin{equation}
    U_1^{(0)}(x) = - \frac{f \beta (2 \beta - 1)^2}{(2 f \beta(\beta-1)+1)^2} \,.
\end{equation}
The finite-$\hbar$ correction $U_1^{(1)}(x)$ is lengthy and may be obtained from ${\cal A}_1$ and ${\cal B}_1$.%
\footnote{
Unlike in the helicity-2 case where $U_2^{(1)}(x)$ was $\w$ and $\q$-independent, the correction $U_1^{(1)}(x)$ depends on the ratio $\w^2/\q^2$ (but not on $\w$ and $\q$ separately).
}
In the $\hbar\to0$ limit, the ground state energy will be given by the minimum of $U_1^{(0)}(x)$. Expanding $U_1^{(0)}(x)$ near the boundary, we find
\begin{equation}
    U_1^{(0)}(x) = -1 + \frac{\beta (3 - 2 \beta)}{(2 \beta-1)^2} x^2 + O(x^4)\,.
\end{equation}
Thus, in the shear channel, a bound state with $E < -1$ (and hence the instability of the boosted black brane) exists for $\beta > \frac{3}{2}$, corresponding to $\lgb < - \frac{3}{4}$.
Again, our potential $U_1^{(0)}(x)$ is precisely the negative of the effective potential that was used in previous discussions of boundary causality violation in GB gravity~\cite{Buchel:2009tt}, and our condition of instability in the limit $\hbar\to0$ is exactly the same as the condition of boundary superluminality in \cite{Buchel:2009tt}. The potential $U_1^{(0)}(x)$ is shown in Figure~\ref{fig:U20_analytic}; the potential $U_1^{(0)}(y)$ looks qualitatively the same, ``stretched'' over the range of $y$. 
The minimum value $E_0(\beta) = {\rm min}\, U_1^{(0)}$ is easily found, 
\begin{align}
  E_0(\beta) = - \frac{(2\beta - 1)^2}{8(\beta - 1)} \,,
\end{align}
and is plotted in Fig.~\ref{fig:hel2_E0}.

\begin{figure}
\centering
\includegraphics[width=0.46\textwidth]{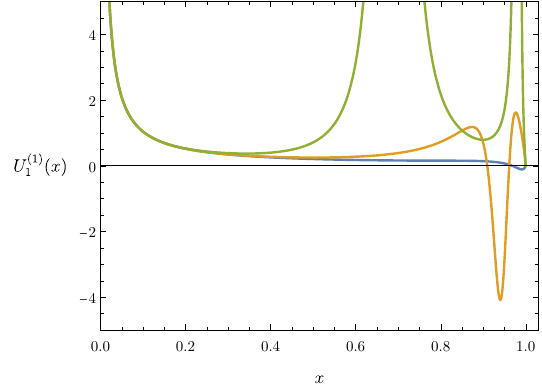}
\hspace{0.05\textwidth}
\includegraphics[width=0.46\textwidth]{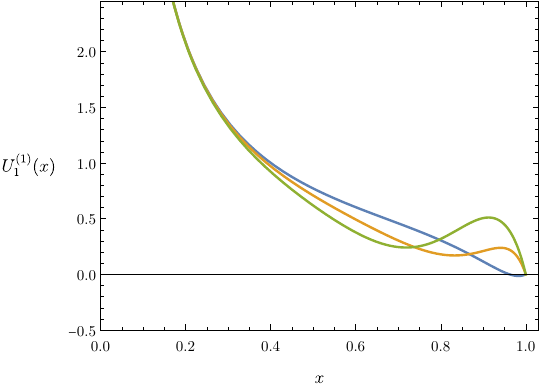}
\caption{
  Left: plots of $U_1^{(1)}(x)$ for $\beta = 3$ (unstable region), and $v = 0.2$ (blue), $v=0.78$ (orange), and $v=0.9$ (green). The potential $U_1^{(1)}(x)$ for $v=0.9$ has two singularities. Right: plots of $U_1^{(1)}(x)$ for $\beta = 1.49$ (stable region), for the same values of $v$.
}
\label{fig:hel1u1}
\end{figure}

The $U_1^{(1)}$ contribution to the potential does not have a definite sign in the shear channel, and depends on both $\beta$ and $v$. Near the boundary, we again have
\begin{align}
  U_1(x) = \frac{15\hbar^2}{16 \beta^2} \frac{1}{x} - 1 + O(\hbar^2 x) \,,
\end{align}
which is the same as in the scalar channel, due to the black brane background being asymptotically AdS.
The near-horizon behaviour is determined by both $\beta$ and $v$. 
For $\beta < 3/2$, the potential $U_1^{(1)}(x)$ is not sign-definite, and can form a dip near the horizon. 
When $\beta>3/2$, for a given $0<v^2<1$, the potential $U_1^{(1)}(x)$ will develop singularities when 
\begin{align}
  \frac{8(\beta - 1)}{(2\beta - 1)^2} \leq v^2 < 1 \,.
\end{align}
The singularities of $U_1(x)$ are double poles with positive residues which are $O(\hbar^2)$; they will not affect the analysis of the ground state in the limit $\hbar\to0$, unless the singularity happens to sit exactly at the minimum of $U_1^{(0)}(x)$. 
Several examples of $U_1^{(1)}(x)$ are shown in Figure~\ref{fig:hel1u1}. 

$U_1^{(0)}(x)$ near the horizon, at $x=1$, behaves as 
\begin{align}
\label{eq:U1h}
 U_1^{(0)}(x)=2\beta (2\beta-1)^2\,\, (x-1)+O((1-x)^2)\,.
\end{align}
Since $\frac{dU_1^{(0)}(x=1)}{dx}>0$, unlike the scalar channel,
the potential never develops a new near-horizon ``dip+bump'' structure, and
there are no additional constraints on $\lgb$
associated with hyperbolicity violation in this channel.

\subsection{Helicity-0 perturbations (sound channel)}
\label{subsec:sound_analytics}
\begin{figure}[t]
\centering
\includegraphics[width=0.6\textwidth]{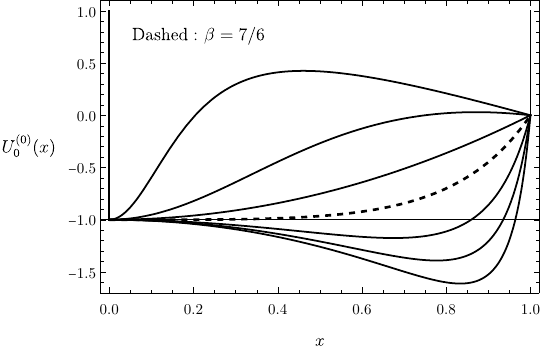}
\caption{
Plots of $U_0^{(0)}(x)$ for $\beta = 0.6, 0.8, 1, 7/6, 1.4, 1.6, 1.8$ in the helicity-0 channel. The potential for $\beta = 7/6$ is shown by the dashed line. The potentials for $\beta < 7/6$ correspond to the curves above the dashed line; the potentials for $\beta > 7/6$ correspond to the curves below the dashed line. Vertical lines represent infinite potential walls, reflecting Dirichlet boundary conditions on $\psi$ at $x=0$ and $x=1$. A bound state in this potential with $E<-1$ implies the existence of an unstable gapped mode in the quasinormal spectrum of the boosted black brane (as $\hbar\to0$) at that value of~$\beta$. Only the potentials with $\beta > 7/6$ (or $\lgb<-7/36$) have such bound states corresponding to unstable modes. 
}
\label{fig:hel0_pot}
\end{figure}

Finally, let us proceed with the analysis of the sound channel. For the equation~\eqref{eq:QNM_master_eq} in the sound channel, the coefficients ${\cal A}_0(x)$ and ${\cal B}_0(x)$, as well as the logarithmic derivative of the integrating factor ${\cal C}_0(x)$, are given in Appendix~\eqref{app:QNM_coeffs_sound}.  The integrating factor once again has the asymptotic behaviour~\eqref{eq:C_asymptotics}, and so we may instead attempt to solve the effective 1D quantum mechanical problem represented by the Schr\"odinger equation for $\psi(y)\equiv \psi_0(y)$,
\begin{equation}
    -\hbar^2 \psi''(y) + (U_0^{(0)}(x) + \hbar^2 U_0^{(1)}(x)) \psi(y) = E \psi(y)\,.
\end{equation}
The $\hbar$-independent part of the potential is given by
\begin{equation}
    U_0^{(0)}(x) = f \beta \frac{4 \beta^2 (\beta-1)^2 f^2+4 f \beta (\beta-1)-8 \beta^2+8 \beta-1}
{(2 f \beta (\beta-1)+1)^2}\,.
\end{equation}
The correction $U_0^{(1)}(x)$ is lengthy and may be obtained from ${\cal A}_0$ and ${\cal B}_0$.%
\footnote{
Unlike in the helicity-2 case where $U_2^{(1)}(x)$ was $\w$ and $\q$-independent, the correction $U_0^{(1)}(x)$ depends on the ratio $\w^2/\q^2$ (but not on $\w$ and $\q$ separately).
}
In the $\hbar\to0$ limit, the ground state energy will be given by the minimum of $U_0^{(0)}(x)$. Expanding $U_0^{(0)}(x)$ near the boundary, we find
\begin{equation}
  \label{eq:u00}
    U_0^{(0)}(x) = -1 + \frac{\beta (7 - 6 \beta)}{(2\beta-1)^2}x^2 + O(x^4)\,.
\end{equation}
Thus, in the sound channel, a bound state with $E < -1$ (and hence the instability of the boosted black brane) exists for $\beta > \frac{7}{6}$, corresponding to $\lgb < - \frac{7}{36}$.
Again, our potential $U_0^{(0)}(x)$ is precisely the negative of the effective potential that was used in previous discussions of boundary causality violation in GB gravity~\cite{Buchel:2009tt}, and our condition of instability in the limit $\hbar\to0$ is exactly the same as the condition of boundary superluminality in \cite{Buchel:2009tt}. The potential $U_0^{(0)}(x)$ is shown in Figure~\ref{fig:hel0_pot}; the potential $U_1^{(0)}(y)$ looks qualitatively the same, ``stretched'' over the range of $y$. 

\begin{figure}
\centering
\includegraphics[width=0.46\textwidth]{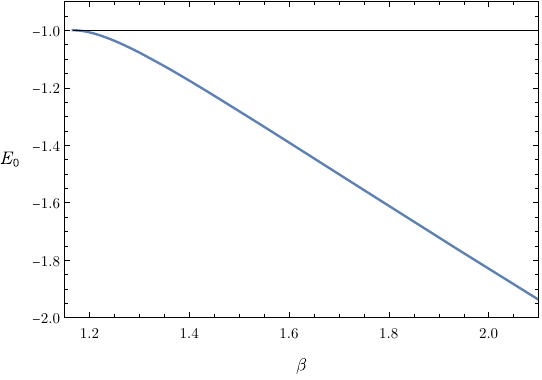}
\hspace{0.05\textwidth}
\includegraphics[width=0.46\textwidth]{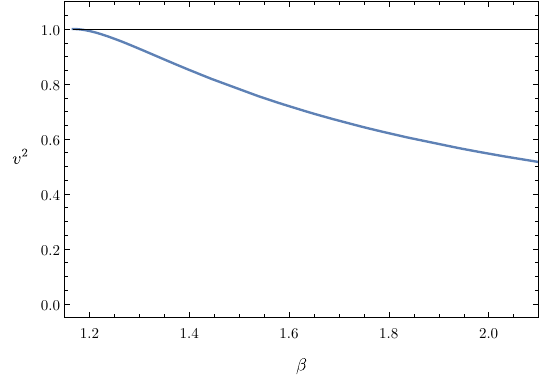}
\caption{
The ground state energy $E_0(\beta)$ in the limit $\hbar\to0$ (left), and the corresponding boost velocity squared $v^2(\beta)$ required to generate an unstable gapped mode at zero wave vector (right) in the sound channel, plotted in the unstable region $\beta > 7/6$. The boost velocity approaches zero at large $\beta$ as $v^2 \sim 1/\beta$.
}
\label{fig:hel0_E0}
\end{figure}

The minimum value of the potential can be found analytically (as $\hbar\to0$), though the expression for $E_0(\beta) = {\rm min}\, U_0^{(0)}$ is not illuminating, and we will not present it here. Near $\beta=7/6$ and $\beta = +\infty$ we have 
\begin{align}
   E_0 = -1 - \frac{81}{10} \left( \beta - \coeff76 \right)^2 + \dots \,,\ \ \ \ 
   E_0 = -\beta + \frac{1}{4\beta} + \frac{1}{4\beta^2} + \dots\,.
\end{align}
The dependence of $E_0$ and of the corresponding $v^2 = -1/E_0$ on $\beta$ is shown in Fig.~\ref{fig:hel0_E0}.

\begin{figure}
\centering
\includegraphics[width=0.46\textwidth]{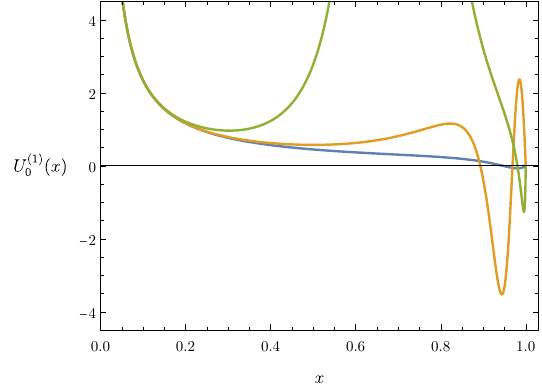}
\hspace{0.05\textwidth}
\includegraphics[width=0.46\textwidth]{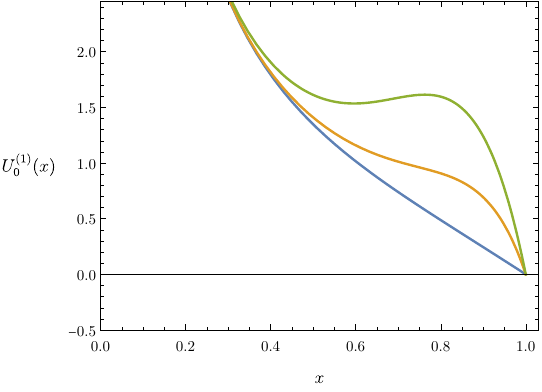}
\caption{
  Left: plots of $U_0^{(1)}(x)$ for $\beta = 2$ (unstable region), and $v = 0.2$ (blue), $v=0.78$ (orange), and $v=0.9$ (green). The potential $U_0^{(1)}(x)$ for $v=0.9$ has one singularity. Right: plots of $U_0^{(1)}(x)$ for $\beta = 1.1$ (stable region), for the same values of $v$.
}
\label{fig:hel0u1}
\end{figure}

The $U_0^{(1)}$ contribution to the potential does not have a definite sign in the sound channel, and depends on both $\beta$ and $v$. Near the boundary, we again have
\begin{align}
  U_0(x) = \frac{15\hbar^2}{16 \beta^2} \frac{1}{x} - 1 + O(\hbar^2 x) \,,
\end{align}
which is the same as in the scalar and shear channels, due to the black brane background being asymptotically AdS.
The near-horizon behaviour is determined by both $\beta$ and $v$. 
When $\beta>7/6$ (unstable region), for a given $0<v^2<1$, the potential $U_0^{(1)}(x)$ will develop singularities when 
\begin{align}
  \frac{48(\beta - 1)}{36\beta^2 - 36\beta + 1} \leq v^2 < 1 \,.
\end{align}
The singularities of $U_0(x)$ are double poles with positive residues which are $O(\hbar^2)$; they will not affect the analysis of the ground state in the limit $\hbar\to0$, unless the singularity happens to sit exactly at the minimum of $U_0^{(0)}(x)$. 
Several examples of $U_0^{(1)}(x)$ are shown in Figure~\ref{fig:hel0u1}. 

$U_0^{(0)}(x)$ near the horizon, at $x=1$, behaves as 
\begin{align}
\label{eq:U0h}
 U_0^{(0)}(x)=2\beta (1-8\beta+8\beta^2)\,\, (x-1)+O((1-x)^2)\,.
\end{align}
Since $U_0^{(0)}(x=0)=-1<0$ \eqref{eq:u00}, as in the scalar channel,
the potential  develops
a ``dip'' and a ``bump'' right in front of the horizon provided
\begin{align}
\label{eq:b2hyp}
1-8\beta+8\beta^2\equiv 1-8\lgb <0\qquad \Longrightarrow\qquad  \lgb>\frac 18\,, 
\end{align}
recovering the remaining interval of $\lgb$
associated with hyperbolicity violation
of GB black branes, see  \eqref{eq:hcl}.

\subsection{Perturbative solutions}
\label{subsec:PT}

We have seen that in the unstable region, the potential $U_a(y)$ in the effective Schr\"odinger equation \eqref{eq:Scheq} has a minimum at $y=y_{\rm min}$, with $U_a(y_{\rm min})<-1$. Expanding the potential about the minimum, we can find the ``energy'' $E=-\w^2/\q^2$ in the small-$\hbar$ expansion, which gives a relation between (imaginary) $\w$ and $\q$ at large $|\q|$. The first correction comes from the ground state energy of the harmonic oscillator. Writing $\w = i\alpha$, $\q = i\kappa$, we have
\begin{align}
\label{eq:E0-QM-1}
  \frac{\alpha^2}{\kappa^2} = |E_0| - \frac{1}{|\kappa|} \left( \coeff12  U_a^{(0)}{}''(y_{\rm min}) \right)^{1/2} + O(1/\kappa^2) \,,
\end{align}
where $E_0$ is the minimum value of $U_a$ as $\hbar\to0$. 
One can find the $O(1/\kappa^2)$ correction by Taylor expanding the potential about $y=y_{\rm min}$. Taking the harmonic oscillator as the starting point, the quartic term in the potential gives $O(1/\kappa^2)$ contribution to the ground state energy in first-order quantum mechanical perturbation theory. The cubic term in the potential gives $O(1/\kappa^2)$ contribution to the ground state energy in second-order perturbation theory. For the potential $U_a = U_a^{(0)} + \hbar^2 U_a^{(1)}$, we find
\begin{align}
\label{eq:E0-QM-2}
  \frac{\alpha^2}{\kappa^2} = |E_0| -  \frac{1}{|\kappa|} \frac{ \sqrt{ U_a^{(0)}{}''} }{\sqrt{2}}  - \frac{1}{\kappa^2} \left( U_a^{(1)} + \frac{1}{16} \frac{U_a^{(0)}{}''''}{U_a^{(0)}{}''} - \frac{11}{144} \frac{ (U_a^{(0)}{}''')^2}{(U_a^{(0)}{}'')^2 } \right) + O(\kappa^{-5/2}) \,.
\end{align}
In the above expression, the potential is a function of $y$, and the $y$-derivatives are evaluated at the minimum of $U_a^{(0)}(y)$. Expressing $\alpha$ as a function of $\kappa$, we have
\begin{align}
  \alpha(\kappa) = \pm |E_0|^{1/2} \kappa \mp \frac{1}{2} \left( \frac{U_a^{(0)}{}''}{2|E_0|} \right)^{\!1/2} + O(1/\kappa)\,,
\end{align}
where the $O(1/\kappa)$ term is straightforwardly obtained from \eqref{eq:E0-QM-2}. We expect the above approximation to work well in the helicity-2 sector where the potential $U_2^{(1)}$ is well-behaved, but not necessarily in helicity-1 and helicity-0 sectors, due to singularities in $U_1^{(1)}(x)$ and $U_0^{(1)}(x)$.

\section{Numerical analysis}
\label{sec:numerics}
In this section we present results of explicit construction of unstable quasinormal modes predicted from analysis in section \ref{sec:analytics}. We will not study these instabilities extensively, rather, we show that they exist; furthermore, we demonstrate how these modes can be obtained from the continuation of {\it stable} black brane quasinormal modes into complex momenta, leading to the violation of the all-reference-frame linearized stability condition~\eqref{linstab},
\begin{equation}
  {\rm Im}(\w) \le |{\rm Im} (\q)| \,.
\label{linstab2}
\end{equation}

\subsection{Unstable modes of boosted black branes}
We will be solving QNM eqs.~\eqref{eq:QNM_master_eq} numerically.  For all helicities $a=0,1,2$, it is convenient to introduce 
\begin{equation}
  Z_a(x)\equiv z_a(x)\  (1-x^2)^{-i\w/2}\ x^2 \,,
\label{strip}
\end{equation}
where we explicitly extracted the incoming wave boundary condition at the horizon, and imposed the normalizabilities of the fluctuations at the asymptotic AdS boundary \cite{Kovtun:2005ev}. As a result, $z_a(x)$ is finite both at the boundary $x=0$, and the GB black brane horizon $x=1$,
\begin{equation}
  z_a(x\to0^+) = z_a^{\rm b} + O(x)\,,\qquad 
  z_a(x\to 1^-) =z_a^{\rm h} + O(1-x) \,.
\label{zbc}
\end{equation}
Since the QNM equations \eqref{eq:QNM_master_eq} are linear, we can always set $z_a^{\rm b}=1$.
The equations for $z_a(x)$ which follow from \eqref{eq:QNM_master_eq} are then solved numerically, with $\w$ and $\q$ taken in \eqref{eq:wq-2}, with fixed real $v$ such that $v^2<1$. The regularity condition as $x\to1^-$ then determines $\Gamma$. We look for solutions where $\Gamma$ is real and positive; based on the quantum mechanical analogy in Sec.~\ref{sec:analytics}, such solutions are expected to exist when $\hbar^2$ is small, that is, when $v^2$ close to $1$, and $\Gamma$ is large. Such solutions can be indeed be found numerically, and we are able to numerically construct the modes which would correspond to ground states in the potential $U_a^{(0)}(x)$.

\begin{figure}
\centering
\includegraphics[width=0.46\textwidth]{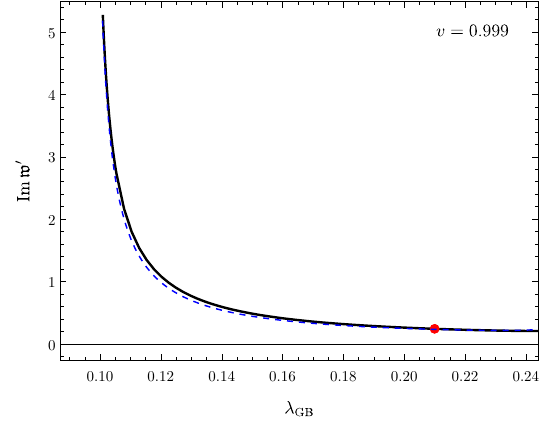}
\hspace{0.05\textwidth}
\includegraphics[width=0.46\textwidth]{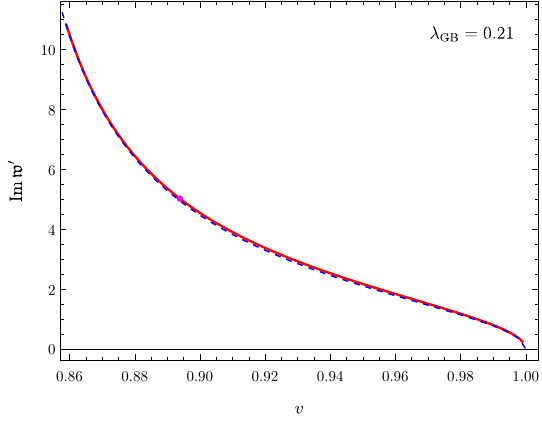}
\caption{
Unstable quasinormal modes of the metric fluctuations in the helicity-2 (scalar) sector. Left panel: The imaginary part $\Gamma$ of the unstable QNM $(\w', \q') = (i\Gamma, 0)$ of the boosted black brane, plotted for boost velocity  $v=0.999$, as a function of Gauss-Bonnet coupling $\lgb$.  The red dot indicates $\lgb = 0.21$. Right panel: the boost velocity dependence of the red dot of the left panel, corresponding to $\lgb=0.21$. The magenta point on the right curve corresponds to the top of the left plot in Fig.~\ref{fig:figure18}, to be discussed later. In both panels, the dashed blue curves show the perturbative approximation to the ground state energy from eq.~\eqref{eq:E0-QM-2}.
}
\label{fig:figure10}
\end{figure}

We start with the scalar channel, solving eq.~\eqref{eq:QNM_master_eq} whose coefficients depend on $\lgb$, as well as on $v$ and $\Gamma$, according to \eqref{eq:wq-2}. For a fixed value of $v$ (close to $1$, as we are looking for unstable modes), the parameters $\Gamma$ and $\lgb$ in the unstable mode will be related; the result of the numerical calculation is shown in Figure~\ref{fig:figure10}, left. Alternatively, for a fixed value of $\lgb$, the parameters $\Gamma$ and $v$ in the unstable mode will be related; the result of the numerical calculation is shown in Figure~\ref{fig:figure10}, right. 
In the left panel, as $\lgb$ decreases, $\Gamma \to +\infty$, and the unstable mode eventually disappears from the spectrum into the complex infinity. It is important to realize that this ``mode removal''  happens before the actual causality bound of $\lgb=9/100$ is reached --- this is so because for a fixed boost velocity $v$ the unstable mode exists only if $\lgb$ is sufficiently far into the causality-violating region, see Fig.~\ref{fig:hel2_min}, right. In the left panel of Fig.~\ref{fig:figure10}, we have chosen $v=0.999$; correspondingly, the range of $\lgb$ when the unstable mode exists is for $\lgb$ between $1/4$ and approximately $0.097$. Similarly, in the right panel of Fig.~\ref{fig:figure10}, the unstable mode only exists if the boost velocity is large enough; for $\lgb=0.21$ ($\beta = 0.7$), the mode disappears into the complex infinity as $v$ approaches approximately $0.828$. 

The disappearance of the unstable mode from the spectrum can also be seen if we look at the radial profile of the bulk solution $z_2(x)$. At fixed $v$, as $\lgb$ approaches a critical value from above, $|z_2^{\rm h}/z_2^{\rm b}|$ diverges, preventing the existence of the solution with the required boundary conditions. Similarly, at fixed $\lgb$, as $v$ approaches a critical value from above, $|z_2^{\rm h}/z_2^{\rm b}|$ diverges, again preventing the existence of the solution with the required boundary conditions. We do not show the corresponding plots, which look qualitatively similar to Fig.~\ref{fig:figure10}.

\begin{figure}
\centering
\includegraphics[width=0.46\textwidth]{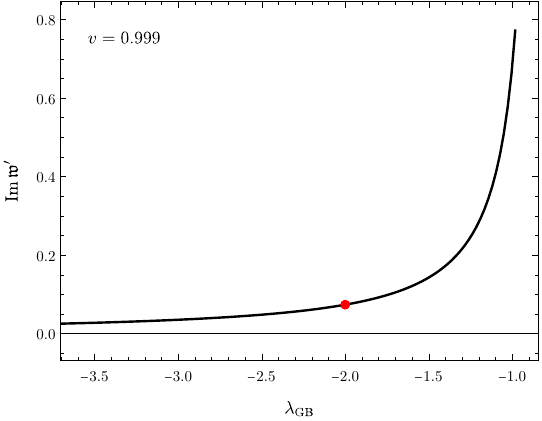}
\hspace{0.05\textwidth}
\includegraphics[width=0.46\textwidth]{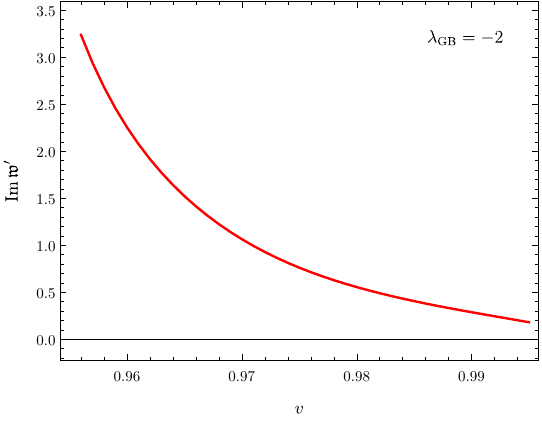}
\caption{
Unstable quasinormal modes of the metric fluctuations in the helicity-1 (shear) sector. Left panel: The imaginary part $\Gamma$ of the unstable QNM $(\w', \q') = (i\Gamma, 0)$ of the boosted black brane, plotted for boost velocity  $v=0.999$, as a function of Gauss-Bonnet coupling $\lgb$.  The red dot indicates $\lgb = -2$. Right panel: the boost velocity dependence of the red dot of the left panel, corresponding to $\lgb=-2$.
}
\label{fig:figure12}
\end{figure}

The story is similar in the shear channel; the unstable mode is shown in Fig.~\ref{fig:figure12}. At fixed $v$, the unstable mode goes off to infinity when $\beta = 1/2 + (1+\sqrt{1-v^2})/v^2$, see Fig.~\ref{fig:hel2_E0}, right. For $v=0.999$ (left panel in Fig.~\ref{fig:figure12}) this gives $\lgb\approx -0.85$, below the causality bound $\lgb=-3/4$. Similarly, at fixed $\lgb$, the unstable mode only exists if the boost velocity exceeds the critical value, which for $\lgb=-2$ (right panel in Fig.~\ref{fig:figure12}) is $v\approx 0.94$. Just like in the scalar channel, the disappearance of the unstable mode from the spectrum can be seen if we look at the radial profile of the bulk solution $z_1(x)$: as $\lgb$ or $v$ approach their respective critical values, $|z_1^{\rm h}/z_1^{\rm b}|$ diverges, preventing the existence of the solution with the required boundary conditions.

\begin{figure}
\centering
\includegraphics[width=0.46\textwidth]{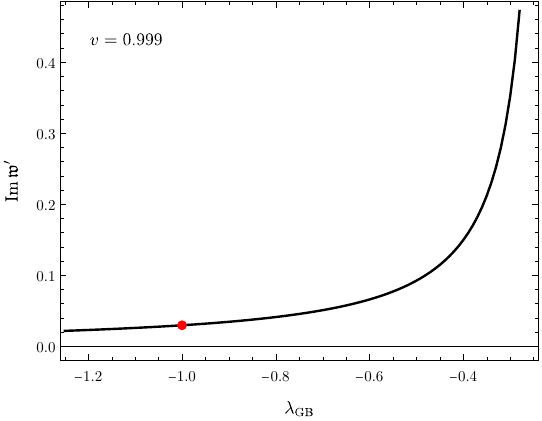}
\hspace{0.05\textwidth}
\includegraphics[width=0.46\textwidth]{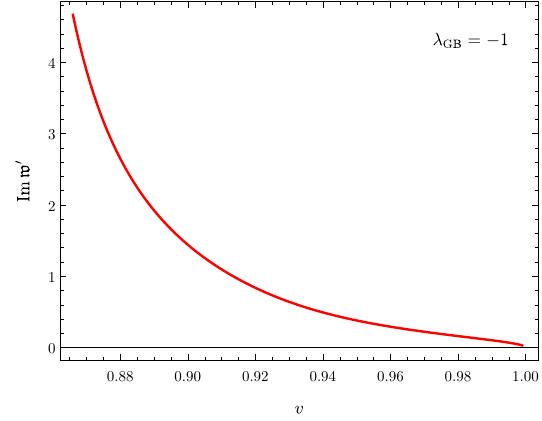}
\caption{
Unstable quasinormal modes of the metric fluctuations in the helicity-0 (sound) sector. Left panel: The imaginary part $\Gamma$ of the unstable QNM $(\w', \q') = (i\Gamma, 0)$ of the boosted black brane, plotted for boost velocity  $v=0.999$, as a function of Gauss-Bonnet coupling $\lgb$.  The red dot indicates $\lgb = -1$. Right panel: the boost velocity dependence of the red dot of the left panel, corresponding to $\lgb=-1$. 
}
\label{fig:figure14}
\end{figure}

The behaviour of the unstable mode is analogous in the sound channel. Near the horizon, the expansion proceeds as
\begin{align}
  z_0(x) = z_0^{\rm h} + \frac{z_0^{\rm h}}{(2 \beta^2 v^2-3)} O(1{-}x)\,,
\end{align}
and the subleading term is singular when $\beta v = \sqrt{3/2}$. This is a spurious singularity, but as it falls into the acausal region of interest for the sound channel for any boost value, it is more convenient to work with the parameter $\tilde{z}_0^{\rm h} \equiv z_0^{\rm h}/(2 \beta^2 v^2-3)$. In fact, every coefficient of the near-horizon expansion is proportional to $\tilde{z}_0^{\rm h}$ rather than $z_0^{\rm h}$. The unstable mode is shown in Fig.~\ref{fig:figure14}. In the left panel (fixed $v$), as $\lgb$ increases towards the causality bound of $-7/36$, the unstable mode eventually moves off to infinity, as $\lgb$ approaches a critical value, which for $v=0.999$ is about $\lgb\approx-0.22$. In the right panel (fixed $\lgb$), the unstable mode disappears from the spectrum as the boost velocity approaches (from above) a critical value, which for $\lgb=-1$ is about $v\approx 0.84$, see Fig.~\ref{fig:hel0_E0}. Analogously to what happens in the scalar and shear channels, one can also see the disappearance of the unstable mode from the spectrum by tracking $|\tilde{z}_0^{\rm h}/z_0^{\rm b}|$ and noticing that the ratio diverges as either $\lgb$ or $v$ approach their respective critical values.

QNMs discussed in Figs.\ref{fig:figure10}-\ref{fig:figure14}
correspond to the ground states of the quantum mechanical potentials discussed
in
sections~\ref{subsec:scalar_analytics}-\ref{subsec:sound_analytics}---
they are the first to become unstable
as $\lgb$ varies outside the causality bounds \eqref{eq:ccl}.
It is straightforward to identify the excited states as well (though we will
not report them here).

\subsection{The complex life of unstable and acausal modes}

Finally, we address the following question: how are the unstable modes we have identified above related to the more familiar quasinormal modes at real~$\q$? In Einstein gravity, the familiar picture of QNM at real $\q$ is the ``Christmas tree'' arrangement in the complex $\w$-plane~\cite{Kovtun:2005ev}. As $|\lgb|$ increases from zero at fixed real $\q$, these modes move in the complex $\w$-plane, plus extra modes with purely imaginary $\w$ come up from $-i\infty$~\cite{Grozdanov:2016vgg}. When $\q$ becomes complex, the QNM move in the complex $\w$-plane, in the process of which ``level-crossings'' are possible, such that two modes which differ at real $\q^2$ swap places as the phase of $\q^2$ changes from $0$ to $2\pi$~\cite{Grozdanov:2019kge, Grozdanov:2019uhi}. In GB gravity, QNMs at complex $\q$ in the shear and sound channels have been explored in~\cite{Grozdanov:2021jfw}, for $|\q|$ up to $O(1)$. In order to study causality violation and instability of black branes in GB gravity, we need $|\q| \gg 1$ in all three channels, which is numerically challenging. At real $\q$, refs.~\cite{Brigante:2007nu, Brigante:2008gz, Buchel:2009tt} gave arguments for the existence of quasinormal modes with $\lim_{\q\to\infty} |{\rm Re}\, \w(\q)/\q| > 1$ when $\lgb$ falls outside the range \eqref{eq:ccl}, however such large-$\q$ ``causality-violating'' modes, to our knowledge, have not been explicitly constructed, either analytically, or numerically. Given the relationship between causality violation (superluminality) and instability discussed in the \nameref{sec:introduction}, it is natural to try to connect the real-$\q$ causality-violating modes of refs.~\cite{Brigante:2007nu, Brigante:2008gz, Buchel:2009tt} to unstable imaginary-$\q$ modes we have discussed in this paper. Such a connection indeed proceeds through complex~$\q$. 

\begin{figure}
\centering
\includegraphics[width=0.46\textwidth]{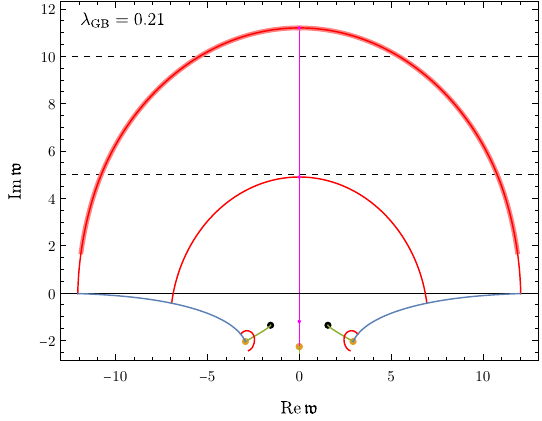}
\hspace{0.05\textwidth}
\includegraphics[width=0.46\textwidth]{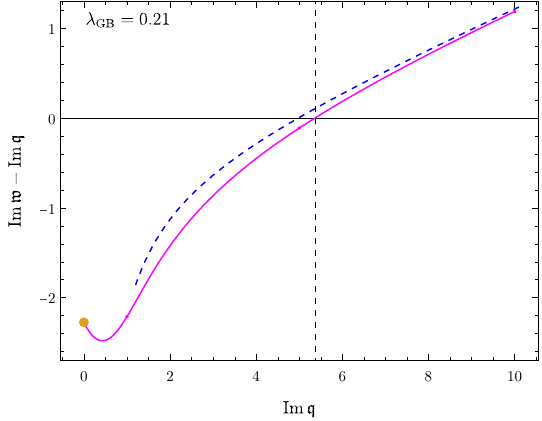}
\caption{
QNMs in the scalar channel. Left panel: We trace $\lgb=0$, $\q=0$ modes of Einstein gravity (the black dots) into the causality violating region $\lgb=0.21$ along the green curves by changing $\lgb$ while keeping $\q=0$. The QNM in the GB theory at $\lgb=0.21$, $\q=0$ are shown by orange dots. Next, the GB QNMs with non-zero real parts are traced along blue curves by changing~$|\q|$ from $0$ to $10$, while keeping $\lgb=0.21$. At $|\q|=\{1,5,10\}$, the phases of $\q$ are rotated from $0$ to $\pi/2$ for the modes on the right, and from $0$ to $-\pi/2$ for the modes on the left. As the phase of $\q$ changes, the QNMs follow the red arcs. Tops of the red arcs at $|\q|=5$ and $|\q|=10$ correspond to purely imaginary $\q$. The thick part of the red arc indicates the modes with $|\q|=10$ for which the covariant stability condition \eqref{linstab2} is violated. The vertical magenta line represents tracing of a purely imaginary QNM (present in the spectrum only for $\lgb\ne 0$) by changing $\q$ from $0$ to $10i$ while keeping $\lgb=0.21$ fixed. Dots on the magenta line correspond to $\q=i$, $\q=5i$, and $\q=10i$ for that mode. Dashed horizontal lines in the left panel are grid lines at ${\rm Im}\,\w=5$, ${\rm Im}\,\w=10$. Right panel: Details of the magenta line in the left panel, showing the trajectory of the GB QNM at $\lgb=0.21$ with ${\rm Re}(\w)=0$, as a function of ${\rm Im}(\q)$. This mode becomes unstable once ${\rm Im}(\q)>5.361(6)$, marked by a vertical black dashed line. The blue dashed line in the right panel shows the perturbative approximation to the ground state ``energy'' in eq.~\eqref{eq:Schrodinger-scalar}, obtained from the first three terms in the $1/|\q|$ expansion of ${\rm Im}\,\w$, following eq.~\eqref{eq:E0-QM-2}.
}
\label{fig:figure16}
\end{figure}

Our results in the scalar channel are presented in Fig.~\ref{fig:figure16}. The figure shows the following. 

{\it a)} We start with the first two scalar-channel QNM of Einstein gravity \cite{Kovtun:2005ev} at $\q=0$, shown by black dots. We then track these modes (green segments) as the Gauss-Bonnet coupling changes from $\lgb = 0$ to $\lgb = 0.21$, still at $\q=0$. As $\lgb$ increases from zero, new modes (absent in Einstein gravity) come up from $-i\infty$ along the imaginary $\w$-axis~\cite{Grozdanov:2016vgg}. At $\lgb=0.21$, the three QNMs at $\q=0$ which are closest to the origin in the complex $\w$-plane are shown by orange dots. 

{\it b)} Next, we track the orange dots as $\q$ changes, while $\lgb$ stays fixed at $\lgb = 0.21$, which is outside the scalar-channel causal range ($\lgb<9/100$). Starting with the right orange dot, we change $\q$ from $0$ to $10$, as the right mode traces the right blue line. Starting with the left orange dot, we change $\q$ from $0$ to $-10$, as the left mode traces the left blue line.%
\footnote{
The differential equation which determines the QNM only depends on $\q^2$, so changing $\q $ from $0$ to $10$ would have produced exactly the same left blue line. 
} 
Starting with the central orange dot, we change $\q$ from $0$ to $10i$ as the central mode traces the magenta line. Small dots on the magenta line correspond to $\q=i$, $\q=5i$, and $\q=10i$. The right panel of Fig.~\ref{fig:figure16} shows details of this magenta trajectory. 

{\it c)} For the left and the right modes which trace blue lines, at fixed $|\q| = \kappa$, we rotate the phase of $\q$ away from zero. As the phase of $\q$ changes, the modes trace red arcs in Fig.~\ref{fig:figure16}, at the corresponding values of $|\q|$. Specifically, for the right mode we take $\q=\kappa e^{i\theta}$, and for the left mode we take $\q = -\kappa e^{-i\theta}$ with $\theta \in [0, \pi/2]$. The arcs shown in Fig.~\ref{fig:figure16} correspond to $\kappa = 1$, $\kappa = 5$, and $\kappa = 10$. At $\theta = \pi/2$, the left and right modes are on the imaginary axis. The thick part of the upper arc denotes the modes with $|\q|=10$ for which the covariant stability condition~\eqref{linstab2} is violated.

We make the following observations. 
\begin{itemize}
\item[{\it i)}]
One can clearly see the violation of the covariant stability condition in the magenta curve in Fig.~\ref{fig:figure16}, as $\q$ crosses a threshold value of approximately $5.36i$. The magenta mode with $|\q|$ larger than that value will make a boosted GB black brane unstable. 

\item[{\it ii)}]
The magenta line in Fig.~\ref{fig:figure16} at large $|\q|$ is the same mode we have plotted earlier in Fig.~\ref{fig:figure10}. This is the mode which corresponds to the ground state at the bottom of the potential $U_2^{(0)}$ in the quantum mechanical picture of Sec.~\ref{subsec:scalar_analytics}. 
The analytic result \eqref{eq:E0-QM-2} for the large-$|\q|$ behaviour of the ground state energy is in excellent agreement with the numerical magenta curve.

\item[{\it iii)}]
We identify the blue lines in Fig.~\ref{fig:figure16} (the lowest-lying QNMs with real $\q$) as the causality-violating modes of refs.~\cite{Brigante:2007nu, Brigante:2008gz}. Our numerical results are consistent with these modes having $|{\rm Re}(\w) /\q | > 1$ and an exponentially small ${\rm Im}(\w)$ when $\q$ is real and $|\q| \to\infty$. 

\item[{\it iv)}]
One can see from Fig.~\ref{fig:figure16} that the causality-violating modes at real $\q$ (blue) and the unstable modes at imaginary $\q$ (magenta) are related by a rotation of the phase of $\q$. 

\item[{\it v)}]
The modes at real $\q$ respect the covariant stability condition \eqref{linstab2}, but violate it once $\q$ is complexified, see thick lines in Fig.~\ref{fig:figure16}. We emphasize that the covariant stability condition \eqref{linstab2} is violated not just on the positive imaginary $\w$-axis, but away from the imaginary $\w$-axis as well.

\item[{\it vi)}]
One can swap ``old'' modes inherited from the ``Christmas tree'' of Einstein gravity with ``new'' modes which are specific to GB. For example, to swap the right orange mode and the central orange mode in Fig.~\ref{fig:figure16}, one can increase real $\q$ from zero, then rotate the phase of $\q$ by $\pi/2$, and then decrease the magnitude $\q$ to zero while keeping $\q$ purely imaginary. Such a swapping by itself is not unexpected~\cite{Grozdanov:2019uhi, Grozdanov:2021jfw}, however the swapping mechanism of Fig.~\ref{fig:figure16} is particularly simple, avoiding complicated mode trajectories. 

\end{itemize}

\begin{figure}
\centering
\includegraphics[width=0.46\textwidth]{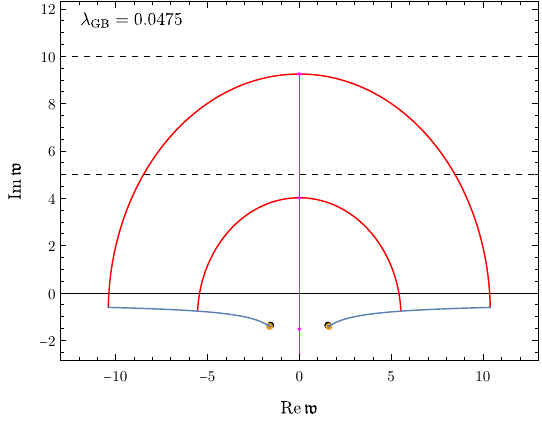}
\hspace{0.05\textwidth}
\includegraphics[width=0.46\textwidth]{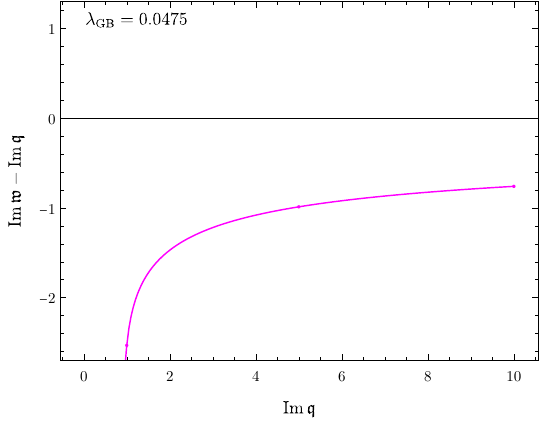}
\caption{
The same modes as in Fig.~\ref{fig:figure16}, but now plotted for $\lgb=0.0475$, which is inside the causal range $\lgb<9/100$. The GB mode on the negative imaginary axis at $\q=0$ is down below, outside the shown plot range. For real $\q$, as $|\q|$ increases, the imaginary parts of the blue modes approach zero, but not exponentially fast like they do in Fig.~\ref{fig:figure16}. The shown red arcs again correspond to $|\q|=1,5,10$, and the magenta dots again to $\q=i, 5i, 10i$. The red arcs at $|\q|=1$ are too small to be visible in the plot. The covariant stability condition \eqref{linstab2} is not violated at complex $\q$ anymore. 
}
\label{fig:figure16a}
\end{figure}

\noindent
It is instructive to compare the modes shown in Fig.~\ref{fig:figure16} at $\lgb > 9/100$ to the same modes for $\lgb < 9/100$, shown in Fig.~\ref{fig:figure16a}. While the covariant stability condition \eqref{linstab2} is violated by the large-$|\q|$ modes shown in Fig.~\ref{fig:figure16}, it is respected by the modes shown in Fig.~\ref{fig:figure16a}.

\section{Discussion}
\label{sec:discussion}

We have shown that black branes in GB gravity \eqref{eq:SGB} are unstable to short-wavelength perturbations unless the GB coupling lies in the range \eqref{eq:ccl}. We have also found the unstable modes numerically. The instability is naively invisible if the background is written in the form \eqref{metric}, however it shows up once a boundary diffeomorphism corresponding to a Lorentz boost is performed. Unsurprisingly, such a pathology is related to the boundary causality violation discussed in refs.~\cite{Brigante:2007nu, Brigante:2008gz, Buchel:2009tt}. We now make several comments. 

\begin{figure}
\centering
\includegraphics[width=0.46\textwidth]{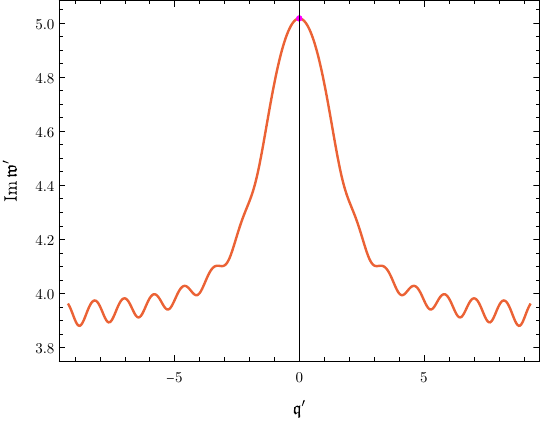}
\hspace{0.05\textwidth}
\includegraphics[width=0.46\textwidth]{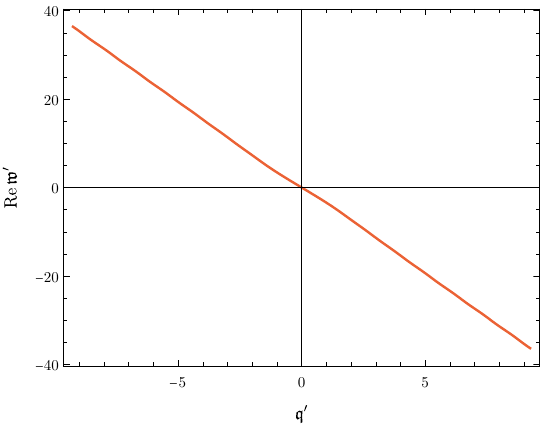}
\caption{
The dispersion relation $\w'(\q')$ of the unstable mode at $\lgb=0.21$ in the boosted reference frame, as a function of real momentum $\q'$, in the scalar channel. The boost velocity is chosen to be $v = |\q|/|\w|$, where $\q = 10i$ and $\w\approx11.1877i$ is the top magenta point of Fig.~\ref{fig:figure16}, which gives $v\approx 0.894$. The mode in the boosted reference frame is maximally unstable at $\q'=0$ (indicated by the magenta dot), and the corresponding value $\Gamma = -i\w'(\q'{=}0) >0$ agrees with Fig.~\ref{fig:figure10} (right) at the corresponding value of $v$.  The mode at $\q'=0$ is the boost of the top magenta point of Fig.~\ref{fig:figure16}. Both ${\rm Re}\,\w'$ and ${\rm Im}\,\w'$ oscillate as functions of $\q'$. 
}
\label{fig:figure18}
\end{figure}

\begin{itemize}
\item
The unstable modes we have identified analytically have purely imaginary~$\w$ and $\q$ in the background \eqref{metric}. These modes violate the covariant stability condition~\eqref{linstab2}. 

\item
Once the quasinormal mode equation is rewritten as the Schr\"odinger's equation, the above unstable modes correspond to true bound states in a confining potential, rather than to quasi-stationary states in an open-ended potential. 

\item
The violation of covariant stability is not restricted to purely imaginary $\q$, and we have numerically constructed the modes where the covariant stability condition \eqref{linstab2} is violated for $\q$ with a non-zero real part, as indicated in Fig.~\ref{fig:figure16}. 

\item
We have numerically constructed the ``causality-violating'' modes with $|{\rm Re}\,\w(\q)/\q| > 1$ at large real $\q$, whose existence was suggested in~\cite{Brigante:2007nu, Brigante:2008gz, Buchel:2009tt}. These modes are stable at real~$\q$, but become unstable as the phase of $\q$ increases, violating the covariant stability condition \eqref{linstab2} at complex~$\q$. 

\item
Our results show that the causality-violating modes of refs.~\cite{Brigante:2007nu, Brigante:2008gz, Buchel:2009tt} are not new modes which suddenly appear in the spectrum once $\lgb$ exceeds $9/100$ (in the scalar channel); they are the familiar ``Christmas-tree'' modes, whose large-$\q$ dispersion relation goes superluminal once $\lgb$ crosses into the acausal range. The same goes for shear and sound channels. 

\item
We have studied QNMs of GB black branes ``at rest'', and have argued that the modes will become unstable upon performing a boost. It is important to point out that when boosting a black brane mode $(\w, \q)$ which violates \eqref{linstab2}, a boost which sets ${\rm Im}\,\q'=0$ does not in general set ${\rm Re}\,\q' = 0$ as well (except for the special modes studied in Sec.~\ref{sec:analytics}). Thus, we do have actual unstable branches $\w'(\q')$ with ${\rm Im}\,\w' > 0$ at real $\q'$, see Fig.~\ref{fig:figure18}.

\item
Some earlier works have interpreted constraints on $\lgb$ in terms of the viscosity bound conjecture proposed in~\cite{Kovtun:2004de}, suggesting that causality in boundary field theory is related to a lower bound on $\eta/s$. We do not find such an interpretation convincing. The stability conditions \eqref{eq:hcl} and \eqref{eq:ccl} are simply indications that the holographic setup involving GB black branes as equilibrium states of a putative dual conformal field theory should be thrown out, unless both \eqref{eq:hcl} and \eqref{eq:ccl} are satisfied. In controlled holographic GB constructions such as those described in~\cite{Kats:2007mq, Buchel:2008vz}, the value of $\lgb$ must be parametrically small, $O(1/N)$ in the $N\to\infty$ limit of the dual gauge theory. 

\item
The plots analogous to Figs.~\ref{fig:figure16} and \ref{fig:figure16a} in the shear and sound channels are qualitatively similar to the scalar channel plots, but involve more modes which clutter the picture, and more challenging numerics at large momentum. When $\lgb$ is outside the causal range for the corresponding symmetry channel, hydrodynamic shear and sound modes become unstable as imaginary $\q$ increases from zero. We do not show the corresponding plots here, planning to return to a more systematic investigation of the interplay between causality and stability involving hydrodynamic modes in the future. 

\item
We have constructed the real-$\q$ causality-violating modes of refs.~\cite{Brigante:2007nu, Brigante:2008gz, Buchel:2009tt} numerically. Analytic control of those modes requires a systematic WKB-like large-$\q$ expansion, which we haven't performed. We plan to return to this in the future.

\item
It would be interesting to understand whether the unstable large-$\q$ modes identified here are related to the nonperturbative contributions to near-lightcone thermal stress-tensor correlators discussed in~\cite{Esper:2023jeq}.

\end{itemize}

\paragraph{Acknowledgements.} 
We would like to thank Sera Cremonini and Sa\v{s}o Grozdanov for valuable discussions.
This work was supported in part by
NSERC of Canada through the Discovery Grants program.
We would like to thank the Simons Center for Geometry and Physics, the Kavli Institute for Theoretical Physics, and the Centro de Ciencias de Benasque Pedro Pascual
where parts of this work were completed. 
This research was supported in part by grant NSF PHY-2309135 to the Kavli Institute for Theoretical Physics (KITP).

\appendix
\section{Black brane thermodynamics}
\label{app:thermodynamics}
The thermal equilibrium state of the boundary field theory associated with black brane~\eqref{metric} is characterized by the temperature $T$, the entropy density $s$, the pressure $P$ and the energy density ${\cal E}$. The entropy is the Bekenstein or the Wald entropy%
\footnote{
  Both are the same for the GB gravity, see \cite{Buchel:2010wf}.
} 
of the background geometry:
\begin{equation}
  s=\frac{2\pi r_h^3}{\ell_P^3}\,.
\label{s}
\end{equation}
The temperature is proportional to the surface gravity $\kappa$ at the horizon,
\begin{equation}
T=\frac{\kappa}{2\pi}=\frac{r_h\beta^{1/2}}{\pi}\ \frac{\sqrt{f_1'f_2'}}{2}\bigg|_{x=1}=\frac{r_h\beta^{1/2}}{\pi}\,.
\label{tk}
\end{equation}
Implementing the standard holographic renormalization for the 1-point function
of the boundary stress-energy tensor \cite{Skenderis:2002wp}, we identify
\begin{equation}
{\cal E}=\frac 34 s \,,\qquad P=\frac 14 s \,,\qquad s=8\pi^2T^3\,
\frac{(3c-a)^4}{(5c-a)^3}\,,
\end{equation}
where we used  \eqref{tk} and \eqref{eq:ac} to arrive at $s(T)$.

\section{QNM equation coefficients}\label{app:QNM_coeffs}
The gauge-invariant combinations $Z_a$ in the different helicity sectors of the metric fluctuations $\xi_{MN}$ are defined as follows~\cite{Kovtun:2005ev,Buchel:2009sk}:
\begin{subequations}
  \begin{align}
{\rm scalar\ channel:}\qquad Z_2 &= \xi_1{}^2\,,
\\
{\rm shear\ channel:}\qquad Z_1 &= \q\,\xi_0{}^1+\w\,\xi_3{}^1\,,
\\
{\rm sound\ channel:}\qquad Z_0 &=
-2\q^2 f\beta\,\xi_0{}^0
+4\w\q\,\xi_0{}^3
+2\w^2\,\xi_3{}^3
+\left(\q^2\beta (f-x f')-\w^2\right)\xi\,,
  \end{align}
\end{subequations}
where $\xi\equiv \xi_1{}^1+\xi_2{}^2$, and spatial indices are raised with the background metric~\eqref{metric}, $\xi_M{}^N\equiv g^{L N}\xi_{M L}$.

\subsection{Helicity-2 (scalar channel)}
\label{app:QNM_coeffs_scalar}
The coefficients of the QNM equation~\eqref{eq:QNM_master_eq} in the scalar channel are given by
\begin{subequations}
    \begin{align}
        {\cal A}_2 &=-\biggl(f (2 f \beta (\beta-1)+1)^2 x\biggr)^{-1}\biggl(
4 f^3 \beta^2 (\beta-1)^2+2 f^2 \beta (\beta-1)-f+2\biggr)\,,\\
{\cal B}_2 &= -\biggl(f^2 (2 f \beta (\beta-1)+1)^2 x\biggr)^{-1}
\biggl(f \beta (12 f^2 \beta^2 (\beta-1)^2+12 f \beta (\beta-1)\\
&-8 \beta (\beta-1)+1) \q^2-\w^2 (2 f \beta (\beta-1)+1)^2\biggr)\,.\nonumber
    \end{align}
\end{subequations}

\subsection{Helicity-1 (shear channel)}
\label{app:QNM_coeffs_shear}
The coefficients of the QNM equation~\eqref{eq:QNM_master_eq} in the shear channel are given by
\begin{subequations}
    \begin{align}
{\cal A}_1(x) &= -\biggl(
16 \beta^4 \w^2 (\beta-1)^4 f^5+24 \beta^3 \w^2 (\beta-1)^3 f^4+8 \beta^2 \w^2 (\beta-1)^2 f^3-\beta (16 \beta^4 \q^2\\
&-32 \beta^3 \q^2-8 \beta^3 \w^2+24 \beta^2 \q^2+16 \beta^2 \w^2-8 \beta \q^2-6 \beta \w^2+\q^2-2 \w^2) f^2+\w^2 (8 \beta^2\nonumber\\&-8 \beta-1) f+2 \w^2\biggr)
\biggl(
f (4 \beta^2 \w^2 (\beta-1)^2 f^2-\beta (4 \beta^2 \q^2-4 \beta \q^2-4 \beta \w^2+\q^2+4 \w^2) f\nonumber\\&+\w^2) (2 f \beta (\beta-1)+1)^2 x\biggr)^{-1}\nonumber\,,\\
{\cal B}_1(x) &= \biggl(
4 \beta^2 \w^2 (\beta-1)^2 f^2-\beta (4 \beta^2 \q^2-4 \beta \q^2-4 \beta \w^2+\q^2+4 \w^2) f
+\w^2\biggr)\\&
\times\biggl((2 f \beta (\beta-1)+1)^2 x f^2\biggr)^{-1}\nonumber\,.
    \end{align}
\end{subequations}

The logarithmic derivative of ${\cal C}_1(x)$ is given by
\begin{equation}
\label{eq:shear_logC2}
\begin{split}
    \frac{d}{dx} \ln {\cal C}_1(x) =& -\frac{1}{4}\biggl(
80 \beta^4 \w^2 (\beta-1)^4 f^4-4 \beta^3 (\beta-1)^2 (12 \beta^2 \q^2-12 \beta \q^2-40 \beta \w^2\\
&+3 \q^2+40 \w^2) f^3-16 \beta^2 (\beta-1) (2 \beta^3 \w^2+4 \beta^2 \q^2-4 \beta^2 \w^2-4 \beta \q^2\\
&-5 \beta \w^2+\q^2+7 \w^2) f^2-\beta (32 \beta^3 \w^2+28 \beta^2 \q^2-64 \beta^2 \w^2-28 \beta \q^2\\&+7 \q^2+32 \w^2) f+16 \beta^3 \q^2-16 \beta^2 \q^2-8 \beta^2 \w^2+4 \beta \q^2+8 \beta \w^2+3 \w^2\biggr)\\
&
\biggl((4 \beta^2 \w^2 (\beta-1)^2 f^2-\beta (4 \beta^2 \q^2-4 \beta \q^2-4 \beta \w^2+\q^2+4 \w^2) f\\
&+\w^2) (2 f \beta (\beta-1)+1)^2 x
\biggr)^{-1}\,.
\end{split}
\end{equation}

\subsection{Helicity-0 (sound channel)}
\label{app:QNM_coeffs_sound}
The coefficients of the QNM equation~\eqref{eq:QNM_master_eq} in the sound channel are given by
\begin{subequations}
\begin{align}
        {\cal A}_0(x) &= \biggl(-8 \beta^4 (\beta-1)^3 (6 \beta \w^2+\q^2-6 \w^2) f^5-24 \beta^3 (\beta-1)^2 (2 \beta^2 \q^2-2 \beta \q^2\\
&+3 \beta \w^2+\q^2-3 \w^2) f^4-8 \beta^2 (\beta-1) (6 \beta^2 \q^2-6 \beta \q^2+3 \beta \w^2+2 \q^2-3 \w^2) f^3\nonumber\\
&+\beta (96 \beta^4 \q^2-192 \beta^3 \q^2-24 \beta^3 \w^2+124 \beta^2 \q^2+48 \beta^2 \w^2-28 \beta \q^2-18 \beta \w^2+3 \q^2\nonumber\\
&-6 \w^2) f^2+(8 \beta^3 \q^2-8 \beta^2 \q^2-24 \beta^2 \w^2-4 \beta \q^2+24 \beta \w^2+3 \w^2) f+4 \beta \q^2-6 \w^2\biggr)\nonumber\\
&\biggl(f x (2 f \beta (\beta-1)+1)^2 (2 \beta^2 (\beta-1) (6 \beta \w^2+\q^2-6 \w^2) f^2-\beta (12 \beta^2 \q^2-12 \beta \q^2\nonumber\\&-12 \beta \w^2+\q^2+12 \w^2) f-2 \beta \q^2+3 \w^2)\biggr)^{-1}\nonumber\,,
\end{align}\begin{align}
{\cal B}_0(x) &=\biggl(
8 \beta^4 \q^2 (\beta-1)^3 (12 \beta^4 \w^2 x+2 \beta^3 \q^2 x-24 \beta^3 \w^2 x-2 \beta^2 \q^2 x+12 \beta^2 \w^2 x
+12 \beta^2\nonumber\\
&-12 \beta+3) f^6-4 \beta^3 (\beta-1)^2 (24 \beta^5 \q^4 x-24 \beta^5 \w^4 x-48 \beta^4 \q^4 x-64 \beta^4 \q^2 \w^2 x\nonumber\\
&+72 \beta^4 \w^4 x
+20 \beta^3 \q^4 x+128 \beta^3 \q^2 \w^2 x-72 \beta^3 \w^4 x+4 \beta^2 \q^4 x-64 \beta^2 \q^2 \w^2 x
+24 \beta^2 \w^4 x\nonumber\\&-44 \beta^2 \q^2+44 \beta \q^2-11 \q^2) f^5-8 \beta^2 (\beta-1) (36 \beta^6 \q^2 \w^2 x+24 \beta^5 \q^4 x
-108 \beta^5 \q^2 \w^2 x\nonumber\\&-30 \beta^5 \w^4 x
-48 \beta^4 \q^4 x+82 \beta^4 \q^2 \w^2 x+90 \beta^4 \w^4 x+25 \beta^3 \q^4 x+16 \beta^3 \q^2 \w^2 x-90 \beta^3 \w^4 x\nonumber\\
&-\beta^2 \q^4 x-26 \beta^2 \q^2 \w^2 x+30 \beta^2 \w^4 x+24 \beta^4 \q^2-48 \beta^3 \q^2
+22 \beta^2 \q^2+2 \beta \q^2-2 \q^2) f^4\nonumber\\&+4 \beta (48 \beta^7 \q^4 x-144 \beta^6 \q^4 x-112 \beta^6 \q^2 \w^2 x
+132 \beta^5 \q^4 x+336 \beta^5 \q^2 \w^2 x+60 \beta^5 \w^4 x\nonumber\\
&-24 \beta^4 \q^4 x-324 \beta^4 \q^2 \w^2 x-180 \beta^4 \w^4 x
-13 \beta^3 \q^4 x+88 \beta^3 \q^2 \w^2 x+180 \beta^3 \w^4 x+\beta^2 \q^4 x\nonumber\\
&+12 \beta^2 \q^2 \w^2 x-60 \beta^2 \w^4 x
-40 \beta^4 \q^2+80 \beta^3 \q^2-54 \beta^2 \q^2+14 \beta \q^2-\q^2) f^3\nonumber\\&+\beta (128 \beta^5 \q^4 x-256 \beta^4 \q^4 x
-240 \beta^4 \q^2 \w^2 x+144 \beta^3 \q^4 x+480 \beta^3 \q^2 \w^2 x+120 \beta^3 \w^4 x\nonumber\\&-16 \beta^2 \q^4 x
-250 \beta^2 \q^2 \w^2 x-240 \beta^2 \w^4 x+96 \beta^4 \q^2+\beta \q^4 x+10 \beta \q^2 \w^2 x+120 \beta \w^4 x\nonumber\\
&-192 \beta^3 \q^2+152 \beta^2 \q^2-56 \beta \q^2+8 \q^2) f^2+2 \beta (8 \beta^3 \q^4 x-8 \beta^2 \q^4 x
-24 \beta^2 \q^2 \w^2 x+\beta \q^4 x
\nonumber\\&+24 \beta \q^2 \w^2 x+15 \beta \w^4 x-2 \q^2 \w^2 x-15 \w^4 x-8 \beta^2 \q^2+8 \beta \q^2-2 \q^2) f
-2 \beta \q^2 \w^2 x\nonumber\\&+3 \w^4 x\biggr)
\biggl((2 \beta^2 (\beta-1) (6 \beta \w^2+\q^2-6 \w^2) f^2-\beta (12 \beta^2 \q^2-12 \beta \q^2-12 \beta \w^2+\q^2\nonumber\\
&+12 \w^2) f-2 \beta \q^2+3 \w^2) (2 f \beta (\beta-1)+1)^3 x^2 f^2\biggr)^{-1}\,.
\end{align}
\end{subequations}
The logarithmic derivative of ${\cal C}_0(x)$ is given by
\begin{align}
\label{eq:logderiv_C3}
    \frac{d}{dx} \ln {\cal C}_0(x) &= -\frac14 \biggl(
40 \beta^4 (\beta-1)^3 (6 \beta \w^2+\q^2-6 \w^2) f^4-4 \beta^3 (\beta-1)^2 (12 \beta^2 \q^2\\
&-12 \beta \q^2-120 \beta \w^2-17 \q^2+120 \w^2) f^3-2 \beta^2 (\beta-1) (48 \beta^3 \w^2+68 \beta^2 \q^2-96 \beta^2 \w^2\nonumber\\
&-68 \beta \q^2-120 \beta \w^2-13 \q^2+168 \w^2) f^2-\beta (96 \beta^4 \q^2-192 \beta^3 \q^2+96 \beta^3 \w^2+244 \beta^2 \q^2\nonumber\\
&-192 \beta^2 \w^2-148 \beta \q^2+11 \q^2+96 \w^2) f+48 \beta^3 \q^2-48 \beta^2 \q^2-24 \beta^2 \w^2+2 \beta \q^2\nonumber\\
&+24 \beta \w^2+9 \w^2\biggr)
\biggl(
x (2 f \beta (\beta-1)+1)^2 (2 \beta^2 (\beta-1) (6 \beta \w^2+\q^2-6 \w^2) f^2\nonumber\\
&-\beta (12 \beta^2 \q^2-12 \beta \q^2-12 \beta \w^2+\q^2+12 \w^2) f-2 \beta \q^2+3 \w^2)
\biggr)^{-1}\nonumber\,.
\end{align}

\bibliographystyle{JHEP}
\bibliography{ads-gb}{}

\end{document}